\documentclass[sigconf]{acmart}

\usepackage{algorithm}
\usepackage{algpseudocode}
\usepackage{appendix}

\usepackage{cclicenses}

\usepackage{graphicx}
\graphicspath{{img/}}

\usepackage{tikz}
\def\checkmark{\tikz\fill[scale=0.4](0,.35) -- (.25,0) -- (1,.7) -- (.25,.15) -- cycle;} 

\newcommand{\Cross}{$\mathbin{\tikz [x=1.4ex,y=1.4ex,line width=.2ex, black] \draw (0,0) -- (1,1) (0,1) -- (1,0);}$}%

\usepackage{subfiles}

\usepackage{caption}
\usepackage{subcaption}

\usepackage{balance}

\usepackage{dblfloatfix}

\usepackage{graphicx}

\AtBeginDocument{%
  \providecommand\BibTeX{{%
    \normalfont B\kern-0.5em{\scshape i\kern-0.25em b}\kern-0.8em\TeX}}}

\copyrightyear{2024}
\acmYear{2024}
\setcopyright{rightsretained}
\acmConference[WiSec '24]{Proceedings of the 17th ACM Conference on Security and Privacy in Wireless and Mobile Networks}{May 27--30, 2024}{Seoul, Republic of Korea}
\acmBooktitle{Proceedings of the 17th ACM Conference on Security and Privacy in Wireless and Mobile Networks (WiSec '24), May 27--30, 2024, Seoul, Republic of Korea}
\acmDOI{10.1145/3643833.3656127}
\acmISBN{979-8-4007-0582-3/24/05}

\begin{document}

\author{Yaqi He}
\email{yhe15@gmu.edu}
\affiliation{
 \institution{Wireless Cyber Center\\George Mason University}
 \streetaddress{4450 Rivanna river}
 \city{Fairfax}
 \state{Virginia}
 \country{USA}
 \postcode{22030}
}

\author{Kai Zeng}
\email{kzeng2@gmu.edu}
\affiliation{
 \institution{Wireless Cyber Center\\George Mason University}
 \streetaddress{4450 Rivanna river}
 \city{Fairfax}
 \state{Virginia}
 \country{USA}
 \postcode{22030}
}

\author{Long Jiao}
\email{ljiao@umassd.edu}
\affiliation{
 \institution{Department of Computer Information Science\\University of Massachusetts Dartmouth}
 \streetaddress{285 Old Westport Rd, North Dartmouth, MA 02747}
 \city{Dartmouth}
 \state{Massachusetts}
 \country{USA}
 \postcode{02748}
}

\author{Brian L. Mark}
\email{bmark@gmu.edu}
\affiliation{
 \institution{Wireless Cyber Center\\George Mason University}
 \streetaddress{4450 Rivanna river}
 \city{Fairfax}
 \state{Virginia}
 \country{USA}
 \postcode{22030}
}

\author{Khaled N. Khasawneh}
\email{kkhasawn@gmu.edu}
\affiliation{
 \institution{Electrical and Computer Engineering Department\\George Mason University}
 \streetaddress{4450 Rivanna river}
 \city{Fairfax}
 \state{Virginia}
 \country{USA}
 \postcode{22030}
}


\title{Swipe2Pair: Secure and Fast In-Band Wireless Device Pairing\\}

\begin{abstract}
Wireless device pairing is a critical security mechanism to bootstrap the secure communication between two devices without a pre-shared secret. 
It has been widely used in many Internet of Things (IoT) applications, such as smart-home and smart-health. 
Most existing device pairing mechanisms are based on out-of-band channels, e.g., extra sensors or hardware, to validate the proximity of pairing devices.
However, out-of-band channels are not universal across all wireless devices, so such a scheme is limited to certain application scenarios or conditions.
On the other hand, in-band channel-based device pairing seeks universal applicability by only relying on wireless interfaces.
Existing in-band channel-based pairing schemes either require multiple antennas separated by a good distance on one pairing device, which is not feasible in certain scenarios, or require users to repeat multiple sweeps, which is not optimal in terms of usability. 

Therefore, an in-band wireless device pairing scheme providing high security while maintaining high usability (simple pairing process and minimal user intervention) is highly desired. 
In this work, we propose an easy-to-use mutual authentication device pairing scheme, named Swipe2Pair, based on the proximity of pairing devices and randomization of wireless transmission power.
We conduct extensive security analysis and collect considerable experimental data under various settings across different environments.
Experimental results show that Swipe2Pair achieves high security and usability.
It only takes less than one second
to complete the pairing process with a simple swipe of one device in front of the other.

\end{abstract}



\maketitle

\vspace{-.2cm}
\section{Introduction}

The widespread adoption of Internet of Things (IoT) technology across multiple fields has ushered in a new era of societal transformation, from commercial sectors to our daily lives.  
IoT systems generate, transmit, and process a vast amount of data, much of which includes sensitive/personal information.
Secure IoT device communication and networking are of utmost importance. 
Inadequate security measures may lead to a range of serious consequences including unauthorized access, privacy breaches, and even threats to life \cite{Yang2017IoTJ}. 
Among various security mechanisms/protocols to safeguard IoT device communication, wireless device pairing has been widely adopted to bootstrap the secure communication between the IoT device and the user device (e.g., smartphone) or access network without a pre-shared secret.  

Existing device pairing protocols can be generally classified into two categories: out-of-band (OOB) channel-based and in-band channel-based. 
OOB channel-based schemes usually rely on auxiliary interfaces and hardware (sensors) to prove/validate the location or context proximity of pairing devices. 
For example, the passkey-based schemes require the auxiliary I/O hardware. QR-code-based schemes require cameras for scanning. 
OOB pairing cannot run on IoT devices without user interfaces and auxiliary hardware/sensors. 
As an alternative, in-band pairing protocols have become popular due to the potential to rely only on wireless interfaces. 
However, the existing in-band based device pairing schemes either require well-separated multiple antennas on one pairing device (i.e., two antennas at two ends on top of the screen of a laptop in the GoodNeighbor scheme \cite{Cai:2011}), which is not always available especially for small-size wireless devices, or require users to repeat multiple
sweeps (i.e., SFIRE in \cite{ghose2020band}), which is not optimal in terms of usability.
Furthermore, GoodNeighbor only provides one-way authentication. 

To the best of our knowledge, in-band device pairing schemes that provide mutual authentication with high usability are missing from the literature. 
To fill this gap, we propose Swipe2Pair, a novel in-band device pairing scheme based on the location proximity of pairing devices and randomized wireless transmission power. 
The advantages of Swipe2Pair are multifold: a) It does not need auxiliary OO\textit{B} channels or sensors. Swipe2Pair can run on IoT devices with standard wireless interfaces (e.g., WiFi, Bluetooth, Zigbee, etc.), which makes it a viable solution for most IoT devices. b) It does not 
involve tedious user interactions or sequential actions and only requires a simple swiping action. Swipe2Pair thus has a short pairing delay (i.e., less than 1 second) and provides better usability than
existing schemes. c) It not only supports mutual authentication but can also defend against powerful attackers with capabilities to accurately infer locations/motions of pairing devices. 
These advantages make Swipe2Pair a robust and usable device pairing solution. 

The main contributions of this work are summarized as follows: 
\begin{itemize}
    \item We propose a new in-band device pairing scheme, Swipe2Pair, which is a universal solution that only utilizes available wireless interfaces on pairing devices.  
    
    \item We adopt transmission power randomization and exploit intrinsic channel fading properties (i.e., a remote attacker experiences higher fading channel variations than closeby pairing devices) to defend against the advanced attackers who may infer the precise distance/location and motion information of pairing devices.

    \item We conduct a thorough security analysis to assess the security strength of Swipe2Pair under different types of attackers, ranging from general attackers to supreme attackers who can accurately infer the location/motion information of pairing devices. 
        
    \item We conduct extensive experiments to validate the security and usability of Swipe2Pair in various environments, demonstrating the robustness of Swipe2Pair.
\end{itemize}

The remainder of the paper is organized as follows. In Section II, we review the state-of-the-art techniques in secure device pairing. In Section III, we describe our system model and attack model. In Section IV, we propose the Swipe2Pair protocol. Section V analyzes the security strength of Swipe2Pair under general, advanced, and supreme attacks. 
Section VI presents the experimental results under different environments. 
Concluding remarks are given in Section VII.

\vspace{-.3cm}
\section{Related work}

Wireless device pairing schemes can be generally classified into two categories: OOB channel-based~\cite{Mayrhofer:2009, wang2015wave, Zhang:2017, Han:2018, Li:2020, jin2014magpairing, yan2019towards} and in-band channel-based~\cite{Cai:2011, ghose2020band}. 

OOB channel-based schemes usually rely on auxiliary hardware or sensors to verify the location or context proximity of the pairing devices. 
Location-proximity based schemes in the OOB channel classify two different categories depending on the physical accessibility of the IoT device.
The inaccessible IoT schemes include Move2Auth~\cite{Zhang:2017}, which
leverages the significant RSSI variation with the inertial accelerometer or gyroscope of the user's smartphone in proximity. Similar scenarios are commonly seen in healthcare involving Implantable Medical Devices (IMDs). 
In these scenarios, extra sensors such as ultrasound transducer ~\cite{siddiqi2021securing} or electrocardiogram heartbeat reader ~\cite{rostami2013heart} are implemented through the touch-to-access principle to communicate and authenticate the inaccessible IMD inside patients' bodies.

The accessible IoT schemes include T2Pair~\cite{Li:2020}, which
involves engaging the button on the IoT device and the smartphone or wristband worn by the user to produce the random pause intervals for authentication. The scheme in \cite{jin2014magpairing} involves a
closely attached magnetometer, while the scheme in \cite{Mayrhofer:2009} involves shaking the accelerometers embedded in two smartphones
together as the shared secret.
Other innovative schemes in this category authenticate pairing devices through physiological values  derived from daily activities of users, transmitted over the wireless body area network (WBAN) \cite{Roeschlin:2018, xu2017gait, pourbemany2022breathe}. These methods offer the advantage of not requiring deliberate pairing gestures, such as pressing, pushing, or rotating.

Additionally, OOB context-proximity schemes serve as distinctive approaches to verify physical proximity, co-location, or shared environmental characteristics of devices within a physically secure boundary that is invisible to outside attackers. These encompass Perceptio~\cite{Han:2018} and NAN~\cite{saloni2016wifi}, which cooperate with the adjacent authenticated devices to validate a new IoT device through their diverse sensors. Moreover, the audio-based scheme \cite{mei2019listen} captures a common audio context through microphones to verify nearby devices. Furthermore, Eolo~\cite{ibrahim2022eolo} leverages barometric pressure to facilitate key verification for closeby devices.

\begin{figure}
  \centering
    \includegraphics[scale=0.35]{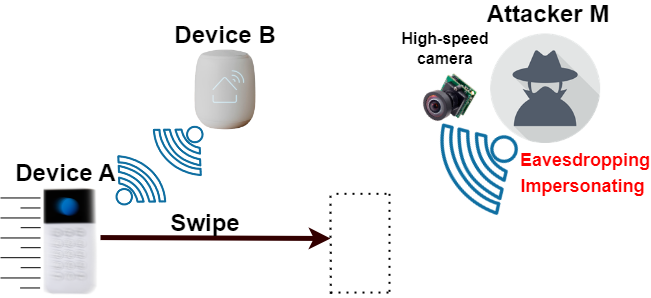}
 \caption{System model comprising adjacent pairing devices \textit{A} and \textit{B}, and a remote attacker \textit{M} who could precisely estimate the location and motion of pairing devices.}
\vspace{-0.6cm}
 \label{fig:system_model}
\end{figure}

\begin{table*}[t]
    \centering
    \begin{tabular}{| p{3cm} | p{1.5cm} | p{1.5cm} | p{3.5cm}  | p{5cm}  | p{1cm}  |}
        \hline
        \textbf{Pairing Approaches} & \textbf{In-band channel} & \textbf{Mutual authen} & \textbf{Sensors/Hardware requirements} & \textbf{User intervention}  & \textbf{Delay}\\
        \hline
        Shake well before use ~\cite{Mayrhofer:2009} & \vfil\hfil \Cross  & \vfil\hfil \checkmark  & Accelerometer  &  hold two devices together and shake & 5s\\
        \hline
        WAVE ~\cite{wang2015wave}& \vfil\hfil \Cross & \vfil\hfil \checkmark & button and LED & wave user arm & 4-6s \\
        \hline
         Move2Auth ~\cite{Zhang:2017} & \vfil\hfil \Cross & \vfil\hfil \Cross & Accelerometer, Gyroscope & Move smartphone back and forth and rotate & 3s\\
         \hline
        Perceptio ~\cite{Han:2018}& \vfil\hfil \Cross & \vfil\hfil \checkmark & Environmental sensors & Null (Auto pairing) & >10s\\
        \hline
        T2pair ~\cite{Li:2020} & \vfil\hfil \Cross & \vfil\hfil \checkmark & Button and smartwatch & Press button multiple times with random pause & 7s \\
         \hline
        Good Neighbour ~\cite{Cai:2011} & \vfil\hfil \checkmark & \vfil\hfil \Cross & At least two antennas on one device & Hold smartphone and move it from one receive antenna to the other & 11.6s\\
        \hline
        SFIRE ~\cite{ghose2020band} & \vfil\hfil \checkmark & \vfil\hfil \checkmark & Helper: Smartphone & Sweep helper at least 4 times (2s/sweep)&  >8s \\
         \hline
         Swipe2Pair (this work) & \vfil\hfil \checkmark & \vfil\hfil \checkmark & No sensor, extra helper, or multiple antennas required & Quick swipe once & $<$1s \\
         \hline
    \end{tabular}
    \caption{Security and Usability Comparison between Representative Device Pairing Schemes}
    \vspace{-0.8cm}
    \label{tab: dp}
\end{table*}

As we can see, OOB channel-based schemes require extra hardware or sensors, which may not be available on all IoT devices.
Meanwhile, heavily relying on the extra hardware and sensors deviates from the purpose of the universal pairing solution.
To achieve better applicability, in-band channel-based solutions are proposed to utilize only wireless interfaces to verify the location proximity of pairing devices.

The in-band proximity-based approaches \cite{khalfaoui2021security} have gained prominence by capitalizing on alternative channels to establish the authenticity of the users and thwart impersonation by Man-in-the-Middle (MitM) attackers. These approaches include 
Good Neighbour~\cite{Cai:2011}, which uses multiple antennas and received signal strength ratio to filter out attackers. However, multiple antennas and an appropriate distance between antennas are not always achievable in all device pairing scenarios.

Other approaches in this category may also exploit in-band channels, exemplified by \cite{ghose2020band}, which relies on the exploitation of physical signal propagation laws or \cite{ghose2017help}, which leverages a helper to detect message manipulation and signal cancellation to resist MitM attacks. However, not every IoT device is accessible or can be authenticated with the assistance of a trustworthy service verifier. Though innovative, the particular requirements of the above device pairing approaches are not universally adaptable to all pairing contexts. 

Besides, usability is another critical but ignored aspect of secure device pairing. The OOB channel-based schemes usually leverage rapid motion, pressing, and rotation multiple times to generate significant RSS variation, which harms simple usage. 
It seems that an in-band scheme with friendly usability and strong security has not been achieved yet. We jointly address these issues in Table~\ref{tab: dp} and then propose our solution.

\vspace{-.3cm}
\section{SYSTEM and ATTACKING MODELS}
\label{sec:sys_mod}
\subsection{System Model}

We assume a general device pairing system model illustrated in Fig. \ref{fig:system_model}, which includes two legitimate pairing devices (\textit{A} and \textit{B}) in close proximity and a remote attacker (\textit{M}). 
At least one of the pairing devices is movable, and we denote it as \textit{A}, which could be a smartphone or an IoT device. 
The other pairing device \textit{B} can either be movable or stationary, such as another IoT device or a wireless router/access point (AP).
Both devices have a wireless interface (e.g., WiFi, Bluetooth, or Zigbee, etc.), allowing them to communicate through the wireless channel. 
They can perform
lightweight computation for encryption and decryption. 

The goal of device \textit{A} and device \textit{B} is to bootstrap a secure channel, i.e., establish a shared secret, through wireless communication without a pre-shared secret or the help of a trusted third party. 
One application example is an IoT device entering a home WiFi network for the first time.
However, the IoT device does not have any input interface, such as a keyboard or touch screen, for the user to type the password.
Then Device \textit{A} is the IoT device and Device \textit{B} is the wireless router under this setting.
Another application example is pairing two IoT devices, such as a smart pen (Device \textit{A}) with a smart notebook or smart board (Device \textit{B}).

\subsection{Attacking Model}
\label{AA}

We consider both passive and active attackers with different attacking capabilities, ranging from general attacks to supreme attacks.

\begin{figure*}[t]
  \centering
    \includegraphics[scale=0.185]{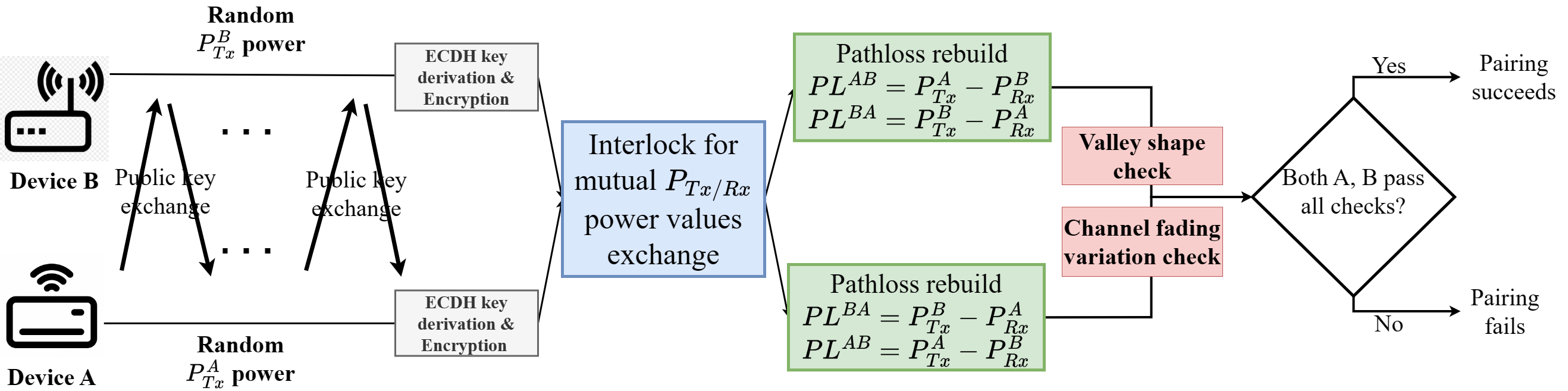}
 \caption{Workflow of Swipe2Pair}
 \label{fig:workflow}
\end{figure*}

\subsubsection{\textbf{Passive attacks}}

Eavesdropping is a typical type of passive attack in which the attacker does not interrupt the device pairing but passively monitors the wireless communications between pairing devices, aiming to steal sensitive information/credentials such as the WiFi password or the secret key established for secure communication.
We assume the attacker can not only overhear all the messages exchanged between the pairing devices but also measure the wireless channel characteristics, such as channel state information or received signal strength (RSS) of the eavesdropping channel.

\subsubsection{\textbf{Active attacks}}
The man-in-the-middle (MitM) attack is considered the most insidious and deceptive active attack against device pairing. The MitM attacker can impersonate both devices by changing its identity to the victims or running an ARP (address resolution protocol) attack. 
In this work, we assume the active attacker can intercept the wireless communication channel of pairing devices and disrupt/overshadow legitimate signals by injecting arbitrary signals with arbitrary power.  

We assume the active attacker can also observe/predict the location/motion of the pairing devices. 
This could be achieved by mounting a high-precision motion camera on the attacking device connected to a powerful computer that can perform real-time computing.
We assume the attacker knows the device pairing protocol.
The attacker cannot be as closeby as the two pairing devices (e.g, 10~cm) to the pairing devices; otherwise, it could be physically detected. The channel characteristics expose the remote location depending on the various circumstances. We will elaborate on this in section~\ref{SEC:PE}.

We consider three types of MitM attackers based on their attacking capabilities: 
\textbf{1) General attacker} who may not know the location or motion of the pairing devices and just follows the pairing protocol.
\textbf{2) Advanced attacker} who can estimate/infer the location/motion of the pairing devices but with a certain degree of estimation error. 
\textbf{3) Supreme attacker} who is assumed to know or precisely estimate the location/motion of the pairing devices. 
We will elaborate on what these different attackers could do and how Swipe2Pair defends against these attacks in the following sections.

\vspace{-.3cm}
\section{PROTOCOL DESIGN}
\label{sec:PD}

In this section, we elaborate on our Swipe2Pair protocol in detail,
as shown in Fig.~\ref{fig:workflow}. The protocol contains four stages: 1) Public key exchange with random transmission (Tx) power and receiving (Rx) power recording; 2) Interlock protocol to exchange encrypted Tx and Rx power information; 3) Computing the bidirectional pathloss between \textit{A} and \textit{B}; and 4) Authentication or proximity validation through valley shape check and channel fading variation check.

\begin{figure}[H]
  \centering
    \includegraphics[scale=0.39]{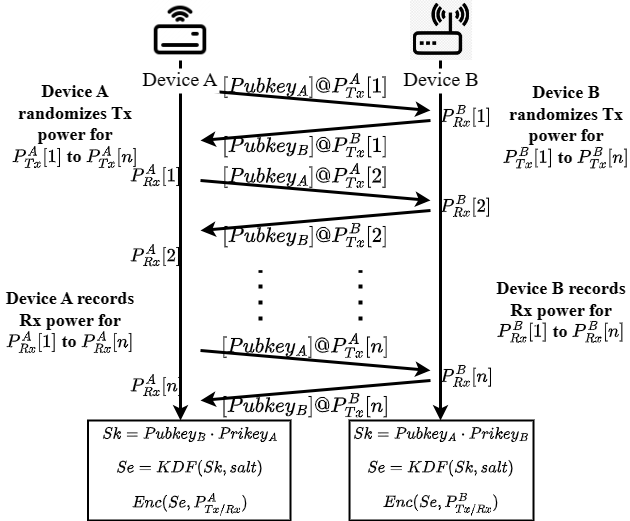}
 \caption{Public key exchange with randomized Tx power and ECDH key derivation 
 }
 \label{fig:pubkey exchange}
\end{figure}

\vspace{-.1cm}

In this stage, the two pairing devices \textit{A} and \textit{B} run the Elliptic Curve Diffie Hellman (ECDH) key exchange protocol by repeatedly exchanging their public keys with random Tx power. 
ECDH is tailored to leverage the cryptographic properties of elliptic curves \cite{vahdati2019comparison} to achieve computational efficiency \cite{kumar2019achieving} and reduced key size \cite{yeh2020energy} while maintaining a comparable level of security. 
These attributes render ECDH particularly well-suited for resource-constrained platforms.
The two pairing devices \textit{A} and \textit{B} follow the following steps:
\begin{enumerate}

    \item Device \textit{A} initializes multiple identical packet Tx, with each packet $i$ ($1\leq i\leq n$) containing \textit{A}'s public key ($Pubkey_{A}$), transmitted at a randomized Tx power $P^{A}_{Tx}[i]$.
    At the same time, the user simply swipes Device \textit{A} from the left side to the right side of Device \textit{B} to accomplish the pairing process. In this process, Device \textit{A} records its random Tx power $P^{A}_{Tx}[i]$. 

    \item On receiving packet $i$ from Device \textit{A}, Device \textit{B} immediately sends back a reply packet containing \textit{B}'s public key ($Pubkey_{B}$)  with a randomized Tx power $P^{B}_{Tx}[i]$. 
    At the same time, \textit{B} records its random Tx power $P^{B}_{Tx}[i]$ and Rx power $P^{B}_{Rx}[i]$ of the received packet from \textit{A}.

    \item When \textit{A} receives the reply packet $i$ from \textit{B}, it records its corresponding Rx power $P^{A}_{Rx}[i]$. 
    
    \item \textit{A} and \textit{B} repeat the above bidirectional public key exchange and channel probing multiple times with a different random Tx power each time until all the $n$ packets have been exchanged. These can be done during the user's fast swipe within 1 second.
    
    \item Both \textit{A} and \textit{B} concatenate their recorded Tx power series and Rx power series together, denoted as \\
    $P^A_{Tx/Rx}= [P^{A}_{Tx}[1], ..., P^{A}_{Tx}[n] | P^{A}_{Rx}[1],..., P^{A}_{Rx}[n]]$ \\ and \\
    $P^B_{Tx/Rx}=[P^{B}_{Tx}[1], ..., P^{B}_{Tx}[n] | P^{B}_{Rx}[1], ..., P^{B}_{Rx}[n]]$, \\respectively.

    \item \textit{A} and \textit{B} derive the shared secret key independently at each side after the user's swiping by multiplying their private key with the exchanged public key received from the other side. Specifically, \textit{A} derives the shared secret key through $ S_k = Pubkey_{B} \cdot Prikey_{A}$ and \textit{B} derives the shared secret key through $S_k = Pubkey_{A} \cdot Prikey_{B}$. Note that the multiplication operation is defined on the elliptic curve over a finite field. After that, an efficient key derivation function (KDF)~\cite{mcginthy2018session} is applied to derive the session key $S_e$ from the shared secret key to secure the power information exchange.

    \item \textit{A} and \textit{B} encrypt their power series information $P^{A}_{Tx/Rx}$ and $P^{B}_{Tx/Rx}$, respectively, with the shared session key $S_e$ and exchange the ciphertexts using the interlock protocol described in the following subsection.
\end{enumerate}

\vspace{-.3cm}
\subsection{Interlock Protocol}
The ciphertext is encrypted with the session key using the AES-ECB block cipher ~\cite{wu2017attack}. Each cipher block is 128 bits. 
The advantage of the interlock protocol \cite{bellovin1994attack} relies on the characteristic that half of an encrypted message cannot be decrypted even with the decryption key. It divides each 128-bit cipher block into two halves and consecutively exchanges the two halves of the cipher blocks between pairing devices. Notice the second half of the cipher block exchange will not start until the first half exchange has been completed from both sides. 

The interlock protocol \cite{rivest1984expose}  ensures non-repudiation for both sides, which prevents \textit{M} from intentionally falsifying the Tx and Rx power information to pass the authentication checks (to be described in Section~\ref{ss:authentication}). Without the interlock protocols, \textit{M} can impersonate \textit{A} or \textit{B} to communicate with the other pairing device and infer the pathloss information to pass all the authentication checks.

\subsection{Pathloss Computing}
\label{ss:pathloss}
After the Tx and Rx power information exchange, both devices \textit{A} and \textit{B} can compute the bidirectional channel pathloss for each channel probing $i$ as follows, assuming a lognormal shadowing pathloss model and considering a measurement error:
\begin{equation} \label{eq:A-B}
\begin{split}
PL^{AB}[i] = P_{Tx}^{\text{A}}[i]- P_{Rx}^{\text{B}}[i]= 10 \alpha  \log_{10}\left (\frac{4\pi ~d_{\text{AB}_{i} }}{\lambda} \right )+\chi_{AB_{i}} + e_{AB_i}\\
PL^{BA}[i]=P_{Tx}^{\text{B}}[i]- P_{Rx}^{\text{A}}[i]= 10  \alpha  \log_{10}\left (\frac{4\pi ~d_{\text{BA}_{i}}} {\lambda} \right )+\chi_{BA_{i}} + e_{BA_i}\\
\end{split}
\end{equation}
where the wavelength $\lambda=c/f=(3\times 10^8/2.4\times 10^9)$ and $\alpha$ is the pathloss exponent and is usually approximated by 2 for line-of-sight and 2.8 for non-line-of-sight scenarios~\cite{wang2012empirical}. 
$\chi_{AB_{i}}$ and $\chi_{BA_{i}}$ represent channel fading, which
follow a Gaussian distribution $\mathcal{N}(0, \sigma^2_{AB})$.
Assuming that channel reciprocity holds, the instance of channel fading $\chi_{AB_{i}}=\chi_{BA_{i}}$ and $d_{AB_{i}}=d_{BA_{i}}$.
While $e_{AB_i}$ and $e_{BA_i}$ are measurement errors modeled as  independently distributed Gaussian variables following $\mathcal{N}(0,\sigma^2_{B})$ and $\mathcal{N}(0,\sigma^2_{A})$ , respectively.
If the measurement errors are small, $PL^{AB}[i]\approx PL^{BA}[i]$.

\vspace{-.1in}
\subsection{Authentication}
\label{ss:authentication}
According to Fig. \ref{fig:workflow}, after both devices \textit{A} and \textit{B} reconstruct their bidirectional pathloss, they can start to authenticate whether the other side is in proximity through two checks: valley shape check, and channel fading variation check.

\subsubsection{\textbf{Valley shape check}}

The valley shape in the rebuild pathloss reflects the user's swiping motion.
The pathloss experiences a decrease and then an increase as \textit{A} is swiped from the left to right of the stationary \textit{B}.
Note that according to the lognormal shadowing model, the pathloss contains two parts: a deterministic part, which is a function of the distance between the two devices and a random channel fading part. 
That is, the valley shape is not smooth but experiences noise, which may cause small local valleys.
To prevent the oscillating noisy data from being identified as the real valley from the user's swipe, we leverage a robust peak-valley detection algorithm refer in \cite{brakel2014} based on the principle of discrepancy. This algorithm constructs a separate moving mean and standard deviation to identify the valley from previous signals without corrupting the signal threshold for further signal data.

In algorithm \ref{alg:valley-detection}, user input parameters contain lag, threshold, and influence. The lag represents the moving window that calculates the mean and standard derivation. The threshold is used to multiply the moving standard deviation, establishing a boundary that measures the deviation of a new data point from the moving mean, thereby facilitating the peak or valley detection. The influence 
represents how much the new signals affect the calculation of the moving mean and moving standard deviation.
\begin{algorithm}
\caption{Robust peak-valley detection algorithm}
\begin{algorithmic}
\State initiate $lag = 100$, $threshold = 4$, $influ = 0.5$
\State signals $\gets$ vector 0,...,0 of length of y;
\State filteredY $\gets$ y(1),...y(lag);
\State stdFilter, avgFilter $\gets$ null;
\State avgFilter(lag), stdFilter(lag) $\gets$ mean( filteredY ), std( filteredY );

\State
  \For{i=$lag$+1,...,t}
    \If{abs(y(i) - avgFilter(i-1)) > $threshold$ * stdFilter(i-1)}
        \If{y(i) > avgFilter(i-1)}
            \State signals(i) $\gets$ +1;
            \Comment{Positive signal}
        \Else
            \State signals(i) $\gets$ -1;
            \Comment{Negative signal}
        \EndIf
            \State filteredY(i) = $influ$*y(i) + (1 - $influ$)*filteredY(i-1);
    \Else
        \State signals(i) $\gets$ 0;
        \Comment{No signal}
        \State filteredY(i) $\gets$ y(i);
    \EndIf
    \State avgFilter(i) $\gets$ mean(filteredY(i-lag+1),...,filteredY(i));
    \State stdFilter(i) $\gets$ std(filteredY(i-lag+1),...,filteredY(i));
  \EndFor
\end{algorithmic}
\label{alg:valley-detection}
\end{algorithm}

If the valley shape has been recognized, we leverage the approach in~\cite{o2022pragmatic} to find the start and end points. This approach fits a model to data by iterative curve fitting and then uses the best-fit model to 
locate the start and end points by interpolation. It locates the start and end points through the cutoff parameter which is a fraction of 
the peak or valley height (usually 1\%). For more complex shapes with multiple peaks or valleys, this approach can combine multiple simple models (e.g., Gaussian) to fit them. Once the start and end points have been identified, the pathloss data range for the valley shape can be extracted for the subsequent channel fading variation checks.

\subsubsection{\textbf{Channel fading variation check}}
\label{sss:var_check}
The second check is the channel fading variation check, which is based on the observation that the channel fading variation experienced by a remote attacker is usually larger than that between the adjacent legitimate pairing devices.
As we will show in Section \ref{sss:supreme_attacker}, a supreme attacker could pass the first check with the knowledge or accurate estimation of the locations/motions of Devices A and B, but will still fail this check.

The channel fading is sensitive to the change of multipath components.
For example, a transmitted signal propagates to the receiver through multiple paths. Each path contributes to a differently delayed, attenuated, and phase-shifted signal. The slight change in certain multipath components may add up to a significant constructive or destructive relative phase of the delayed signals, which leads to considerable fluctuations in the receiving power.
The multipath-rich indoor environment also leads the receiving power to fluctuate on the order of signal wavelength and contributes to large shadowing variance ~\cite{yang2013rssi}.

It has been observed from several empirical studies that the longer the distance between the transmitter and receiver, the larger the channel fading variation~\cite{shang2014location, adewumi2013rssi, zhu2015improved}. Consequently, the remote \textit{M} should experience more pronounced fluctuations caused by both multipath fading from the indoor environment and shadowing effects from environmental obstacles than the adjacent pairing
devices \textit{A} and \textit{B}. We leverage this phenomenon~\cite{mohammed2016distance,4563424,6380497} to filter the supreme attacker \textit{M} who could pass the valley shape check.

In summary, the Swipe2Pair protocol authentication succeeds only if both pairing devices pass two checks in the authentication. Otherwise, the Swipe2Pair process aborts if any check from any device fails and the two pairing devices will be reset.
We will analyze the security performance of Swipe2Pair in more detail in Sections~\ref{sec:SA} and~\ref{SEC:PE}.

\vspace{-.3cm}
\section{SECURITY ANALYSIS}
\label{sec:SA}

In this section, we analyze the security strength of Swipe2Pair against general, advanced, and supreme attacks, respectively.

\vspace{-.3cm}
\subsection{\textbf{Attacking Strategies}}
Before we start the security analysis, let us examine at what stage and how a MitM attacker could launch the attack. 
We assume the MitM attacker could compromise the communication between Devices A and B in the very beginning of the pairing process.
That is, from \textit{A} and \textit{B}'s perspective, they are each running the pairing protocol with \textit{M}.

Although \textit{M} has no control of or knows \textit{A} or \textit{B}'s randomized Tx power, he can fully control his Tx power and can falsify both his Tx power and Rx power information during the Tx/Rx power information exchange stage.
Note that \textit{M} can generate a unique shared key with \textit{A} ($Sk_{AM}$) and \textit{B} ($Sk_{BM}$), and therefore encrypt the corresponding falsified Tx/Rx power information respectively. 
\textit{A} and \textit{B} will decrypt this information and use the information to reconstruct the pathloss and run the two checks described in the previous section.
The valley shape in the pathloss reconstruction varies a lot depending on the user's swipe technique. This may somehow affect the valley shape check as a usability problem. We will elaborate on its robustness in section \ref{SEC:PE}. 

In the following analysis, we will show that a general attacker is unlikely to pass either the valley shape check or the channel fading variation check.
The advanced and supreme attackers could pass the valley shape check, but not the channel fading variation check.

\vspace{-.35cm}
\subsection{\textbf{Security Analysis Against Attackers with Different Capabilities}}
A critical factor that significantly affects the security performance is \textit{M's} capability to
obtain accurate instantaneous estimates of the
distance $d_{AB_{i}}$ while \textit{A} is moving
during the pairing process. Meanwhile, the pathloss exponent $\alpha$ is considered known to \textit{M} since \textit{M} is in a similar environment with \textit{A} and \textit{B}. 
We categorize \textit{M} into three types based on their capability, as discussed in Section~\ref{sec:sys_mod}: General \textit{M}, Advanced \textit{M}, and Supreme \textit{M}.

\subsubsection{\textbf{Security Analysis Against General Attacker}}
Firstly, General \textit{M} does not use any methods to estimate/infer \textit{A/B}'s location information and just follows the pairing protocol without falsifying Tx/Rx information. 
So the rebuilt pathlosses at \textit{A} and \textit{B} are unlikely to pass the valley shape check based on the Tx/Rx information provided by General \textit{M}. 

\noindent
\textbf{Pathlosses Rebuilt at \textit{A} and \textit{B} under General Attack:}
During the pathloss rebuild process, \textit{M} can impersonate either \textit{A} or \textit{B} to communicate with the other.
Similar to the modeling in Section \ref{ss:pathloss}, the eavesdropping/attacking channel between \textit{A}/\textit{B} and \textit{M} is assumed to follow lognormal shadowing, with a channel fading variation $\chi_{MA}$/$\chi_{MB}$ following $\mathcal{N}(0, \sigma^2_{MA/MB})$ and measurement error $e_{MA}$/$e_{MB}$ following $\mathcal{N}(0, \sigma^2_{A/B})$.
When \textit{M} impersonates \textit{B}, the bidirectional posslosses rebuilt at \textit{A} are:  
\begin{equation} \label{eq:A-M}
\begin{split}
&PL^{AM}[i] = P_{Tx}^{\text{A}}[i]- P_{Rx}^{\text{M}}[i]= 10 \alpha  \log_{10}\left (\frac{4\pi d_{\text{AM}_{i} }}{\lambda} \right )+\chi_{AM_{i}} + e_{AM_{i}}\\
&PL^{MA}[i]=P_{Tx}^{\text{M}}[i]- P_{Rx}^{\text{A}}[i]= 10  \alpha  \log_{10}\left (\frac{4\pi d_{\text{MA}_{i}}} {\lambda} \right )+\chi_{MA_{i}} + e_{MA_{i}}.
\end{split}
\end{equation}
Similarly, when \textit{M} impersonates \textit{A}, the bidirectional pathlosses rebuilt at \textit{B} are:
\begin{equation} \label{eq:M-B}
\begin{split}
&PL^{BM}[i] = P_{Tx}^{\text{B}}[i]- P_{Rx}^{\text{M}}[i]= 10 \alpha  \log_{10}\left (\frac{4\pi d_{\text{BM}_{i} }}{\lambda} \right )+\chi_{BM_{i}} + e_{BM_{i}}\\
&PL^{MB}[i]=P_{Tx}^{\text{M}}[i]- P_{Rx}^{\text{B}}[i]= 10  \alpha  \log_{10}\left (\frac{4\pi d_{\text{MB}_{i}}} {\lambda} \right )+\chi_{MB_{i}} + e_{MB_{i}} .
\end{split}
\end{equation}

\begin{figure*}[!b]

  \centering
    \begin{subfigure}{0.25\textwidth}
        \includegraphics[width=\textwidth]{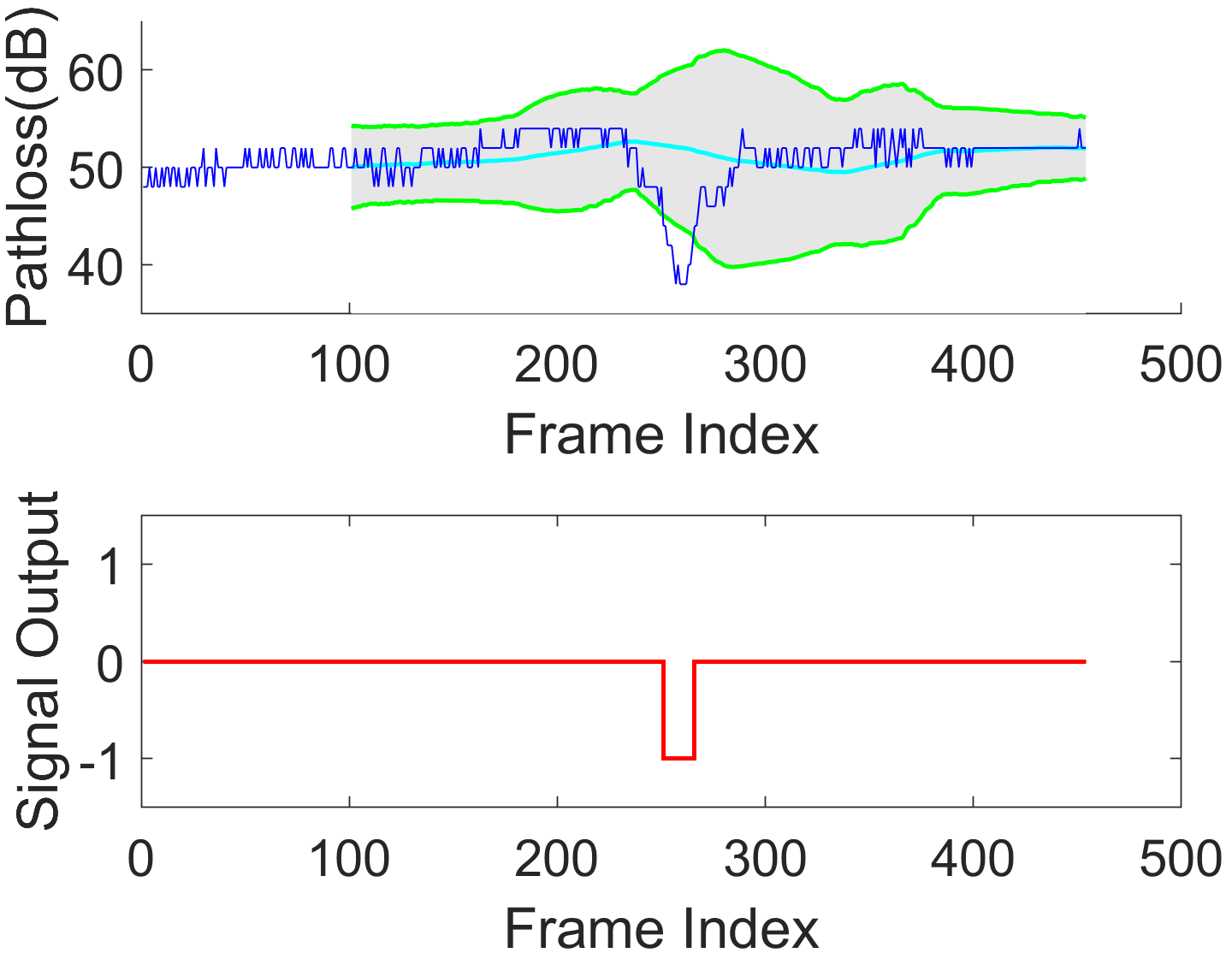}
        \caption{Valley shape detection from pathloss data}
        \label{fig:VS-detect}
    \end{subfigure}
    \hfill
    \begin{subfigure}{0.24\textwidth}
        \includegraphics[width=\textwidth]{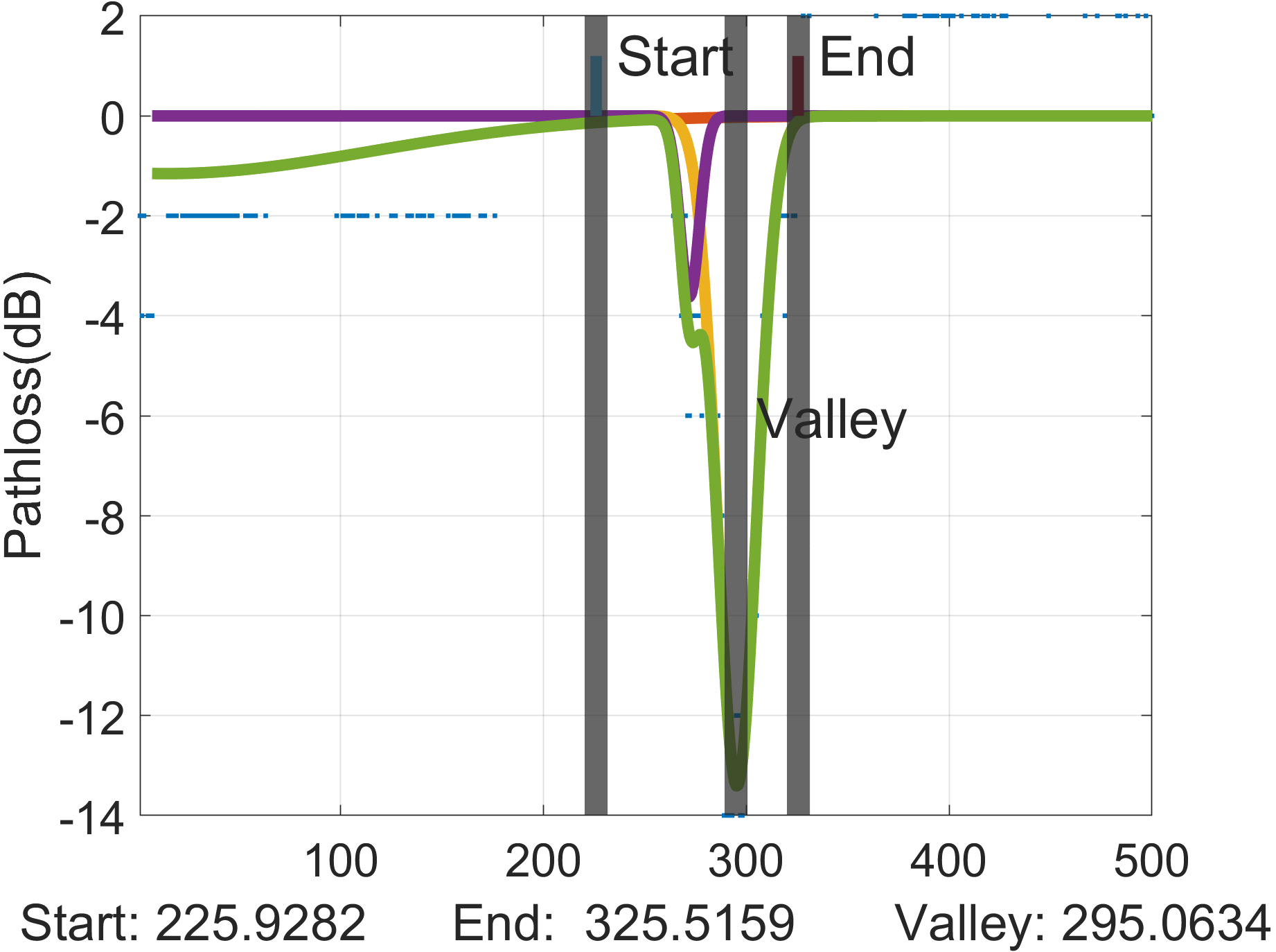}
        \centering
        \caption{Start, end, and valley points identification}
        \label{fig:VS-findpoints}
    \end{subfigure}
    \hfill
        \begin{subfigure}{0.24\textwidth}
        \includegraphics[width=\textwidth]{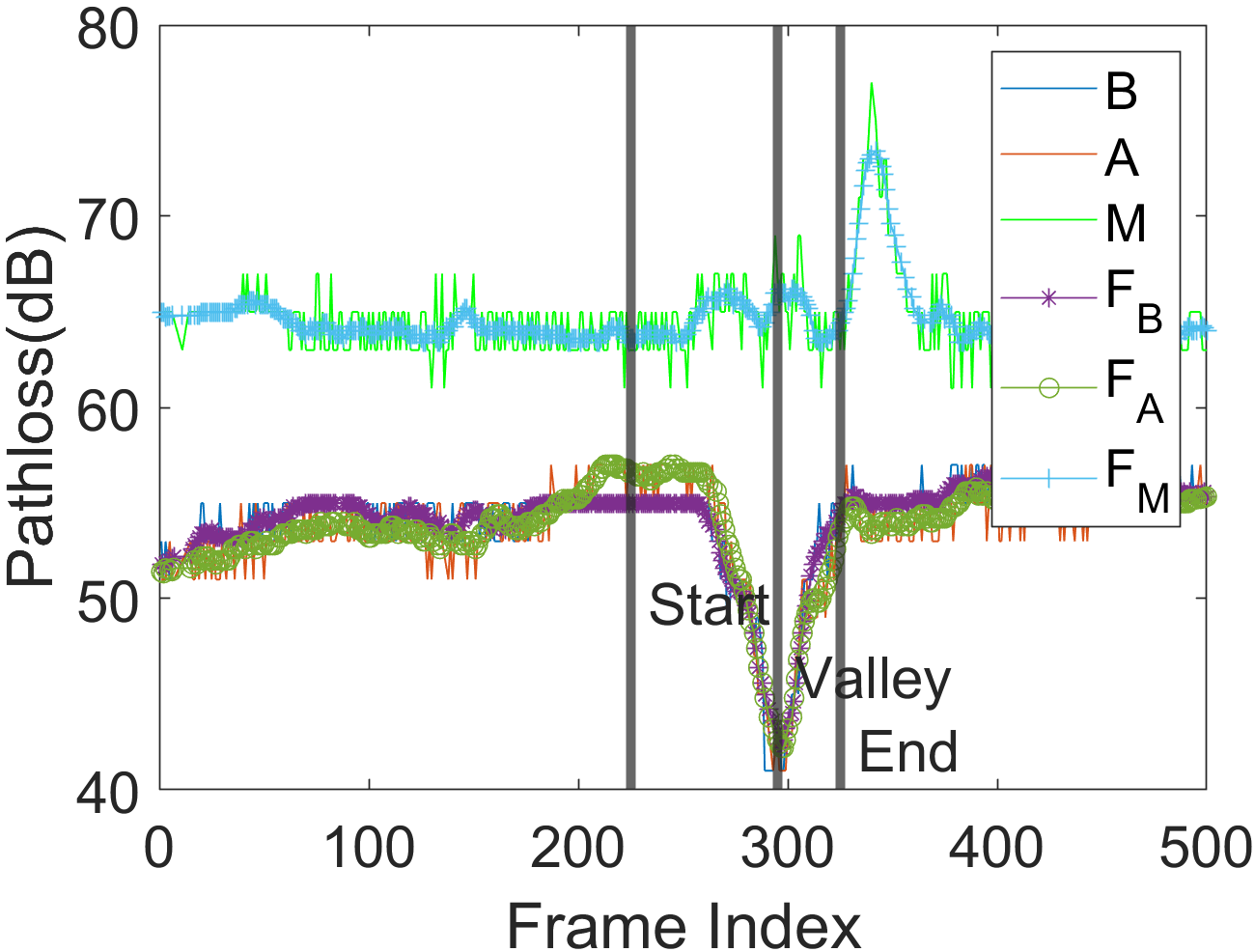}
        \centering
        \caption{Identified points in original pathloss data}
        \label{fig:VS-cutpoints}
    \end{subfigure}
    \hfill
    \begin{subfigure}{0.24\textwidth}
        \includegraphics[width=\textwidth]{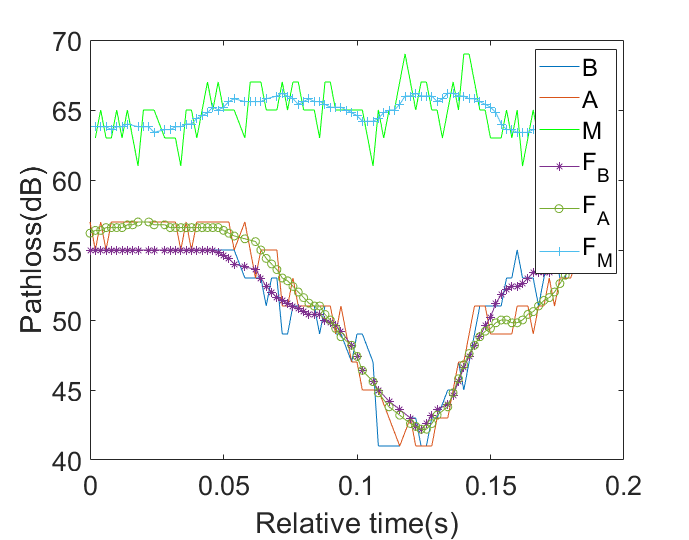}
        \centering
        \caption{Extracted valley shape}
        \label{fig:VS-rangedata}
    \end{subfigure}
    
  \caption{ From the original pathloss data computed through the Tx and Rx powers, the robust peak-valley detection algorithm is implemented to detect a valley shape from noisy data. If the valley shape is detected, we search for the start and end points of the valley shape and extract the valley shape data for security checks.}
 \label{fig:VS}
\end{figure*}

However, since \textit{M} is farther away from \textit{A}/\textit{B} compared to the distance between \textit{A} and \textit{B}, neither \textit{A} nor \textit{B} is likely to observe a deep valley shape.
Since \textit{A} is moving, depending on the relative motion between \textit{A} and \textit{M}, \textit{A} could observe a valley shape, but it is likely to be shallow. 
As we will show in Section ~\ref{SEC:PE} Fig.~\ref{fig:VS-rangedata}, in practice, the valley depth corresponding to the legitimate channel between \textit{A} and \textit{B} is about 15dB with a peak point around 55dB and the valley point around 40 dB. 
While for the attacking channels \textit{A/B-M}, the valley shape is either not observable or is shallow with higher peak and valley points, e.g., around 65dB.
So setting the valley peak around 55dB and valley point around 40dB and a valley width of about 0.2s is sufficient to differentiate an adjacent pairing device from a remote General \textit{M}.

From Eq.~\eqref{eq:A-M}, we can see that both the channel fading variation and measurement error contribute to the observed channel variations.
Therefore, after fully removing the deterministic pathloss components corresponding to distance, the channel variations observed at \textit{A} under attack should follow $\mathcal{N} (0, \sigma_{MA}^2 + \sigma_{A}^2)$.
While from Eq.~\eqref{eq:A-B}, the channel variations observed at \textit{A} without attack should follow $\mathcal{N} (0, \sigma_{BA}^2 + \sigma_{A}^2)$.
As we discussed in Section \ref{sss:var_check}, a farther away attacker usually experiences larger channel fading than adjacent pairing devices.
That is, $\sigma_{MA}>\sigma_{BA}$, so $\sigma_{MA}^2+\sigma_{A}^2 > \sigma_{BA}^2+\sigma_{A}^2$.
We will show in Section \ref{SEC:PE} Fig.~\ref{fig:dist-channel fading} that this principle is validated through our experiments as well.
By setting up a proper threshold (e.g., around 1.27 observed in our experiments) on the sampled standard deviation of the measured channel variations, we can detect the remote attacker.

\subsubsection{\textbf{Security Analysis Against Advanced Attacker}}
We assume Advanced \textit{M} has enhanced capabilities compared to General \textit{M} through observing or guessing a less accurate estimate of the 
instantaneous distances between itself and Device \textit{A}/\textit{B} ($d_{MA_i/MB_i}$) and the distance between \textit{A} and \textit{B} ($d_{{AB}_{i}}$) during the pairing process. 
With the estimate of these distances and the knowledge of the pathloss exponent in the environment, Advanced \textit{M} can falsify its claimed Tx and Rx power information during the Tx/Rx power exchange stage to try to make Device \textit{A/B} believe it is adjacent.
That is, Advanced \textit{M} aims to make the pathloss computed at \textit{A/B} similar to the expected one between \textit{A} and \textit{B} by manipulating the claimed Tx and Rx information to \textit{A/B}.
In the following analysis, we will use the attack on Device \textit{A} as an example. 
A similar analysis can be derived for the attack on Device \textit{B}.

Note that Advanced \textit{M} cannot control the channel fading variation, which is an intrinsic property of the fading channel, nor the measurement error.
What Advanced \textit{M} could achieve is to largely compensate the difference between the pathlosses in channels $A-B$ and $A-M$ by falsifying the claimed Tx and Rx power information, denoted as $P'^{M}_{Tx}$ and $P'^{M}_{Rx}$, respectively.
Considering the distance estimation error at \textit{M}, we can model the estimated distances $\tilde{d}_{AM_i}$ and $\tilde{d}_{AB_i}$ as
\begin{equation} \label{eq:est}
 \tilde{d}_{AM_i}= d_{AM_i} \cdot e^{r_{AM_i}}, ~~   \tilde{d}_{AB_i}= d_{AB_i} \cdot e^{r_{AB_i}},
\end{equation}
where $r_{AM_i}$ and $r_{AB_i}$ are relative estimation errors and can be modeled as Gaussian random variables following $\mathcal{N}(0, \sigma^2_{d})$. 

In order to make the pathloss computed at \textit{A} similar to that between \textit{A} and \textit{B}, Advanced \textit{M} sends the following falsified Tx/Rx power information ($P'^{M}_{Rx}[i]$ and $P'^{M}_{Tx}[i]$) to \textit{A}:
\begin{equation}
\label{e:P'^{M}_{Rx/Tx}}
    \begin{split}
      P'^{M}_{Rx}[i] = P^{M}_{Rx}[i] + 10\alpha log_{10}\left( \frac{\tilde{d}_{AM_i}}{\tilde{d}_{AB_i}}\right)\\
      P'^{M}_{Tx}[i] = P^{M}_{Tx}[i] - 10\alpha log_{10}\left( \frac{\tilde{d}_{AM_i}}{\tilde{d}_{AB_i}}\right)
    \end{split}
\end{equation}

Then the computed  pathloss $PL'^{AM}[i]$ at \textit{A} is:
\begin{equation} \label{e:PL'^{AM}}
\begin{split}
&PL'^{AM}[i] = P_{Tx}^{\text{A}}[i] - P_{Rx}^{'\text{M}}[i] 
= P_{Tx}^{\text{A}}[i] - \left [ P_{Rx}^{\text{M}}[i] + 10\alpha \log_{10} \left( \frac{\tilde{d}_{AM_i}}{\tilde{d}_{AB_i}} \right) \right ] \\
&= 10\alpha \log_{10} \left ( \frac{4 \pi d_{AB_{i}} } {\lambda} \right ) + \chi_{AM_{i}}  + \frac{10\alpha \cdot (r_{AB_{i}}-r_{AM_{i}})} {ln(10)} + e_{AM_{i}}.
\end{split}
\end{equation}

Similarly, we can derive  the computed pathloss 
$PL'^{MA}[i]$ as:
\begin{equation} \label{e:PL'^{MA}}
\begin{split}
&PL'^{MA}[i] = P_{Tx}^{'\text{M}}[i] - P_{Rx}^{\text{A}}[i] 
= \left [ P_{Tx}^{\text{M}}[i] - 10\alpha \log_{10} \left( \frac{\tilde{d}_{AM_i}}{\tilde{d}_{AB_i}} \right) \right ] - P_{Rx}^{\text{A}}[i]  \\
&= 10\alpha \log_{10} \left ( \frac{4 \pi d_{AB_{i}} } {\lambda} \right ) + \chi_{MA_{i}}  + \frac{10\alpha \cdot (r_{AB_{i}}-r_{AM_{i}})} {ln(10)} +  e_{MA_{i}}.
\end{split}
\end{equation}

From Eqs.~\eqref{e:PL'^{AM}} and \eqref{e:PL'^{MA}}, we can see that the distance-dependent deterministic part, $10\alpha \log_{10} \left ( \frac{4 \pi d_{AB_{i}} } {\lambda} \right ) $, in the rebuilt pathloss at \textit{A} equals to that between \textit{A} and \textit{B}.
Therefore, Advanced \textit{M} could pass the valley shape check.

Furthermore, although the extra variation, $\frac{10\alpha \cdot (r_{AB_{i}}-r_{AM_{i}})} {ln(10)}$, is introduced by the distance estimation errors, it is the same in Eqs.~\eqref{e:PL'^{AM}} and \eqref{e:PL'^{MA}}.

However, Advanced \textit{M} is unlikely to pass the channel variation check since the channel variation is increased due to the distance estimation errors and the larger channel fading variation between \textit{A} and \textit{M}.

\subsubsection{\textbf{Security Analysis Against Supreme Attacker}}
\label{sss:supreme_attacker}
Supreme \textit{M} is a special case of Advanced \textit{M}.
We assume the supreme \textit{M} can gain the precise instantaneous distance $d_{AM_{i}}$ and $d_{AB_i}$ without any estimation error.
That is, 
$r_{AB_i}=r_{AM_i}=0$, so $\tilde{d}_{AM_i}=d_{AM_i}$ and $\tilde{d}_{AB_i}=d_{AB_i}$.
We also assume Supreme \textit{M} does not have any measurement error, i.e., $e_{AM_i}=0$.
Therefore, Eq.~\eqref{e:PL'^{AM}} becomes
\begin{equation} \label{e:sup_PL^AM}
\begin{split}
PL'^{AM}[i] = 10\alpha \log_{10} \left ( \frac{4 \pi d_{AB_{i}} } {\lambda} \right ) + \chi_{AM_{i}}.
\end{split}
\end{equation}

Similarly, Eq.~\eqref{e:PL'^{MA}}
becomes
\begin{equation} \label{e:sup_PL^MA}
\begin{split}
PL'^{MA}[i] = 10\alpha \log_{10} \left ( \frac{4 \pi d_{AB_{i}} } {\lambda} \right ) + \chi_{MA_{i}} + e_{MA_{i}}.
\end{split}
\end{equation}
Note that \textit{A} may still experience the measurement error.

Similar to the security analysis against Advanced \textit{M}, we can conclude that Supreme \textit{M} could pass the valley shape check.
Although there is no distance measurement error introduced variation, the variation in the rebuilt pathloss is introduced by the channel fading between \textit{A} and \textit{M}.
As long as $\sigma_{MA}>\sigma_{BA}$, Supreme \textit{M} is likely to fail the channel fading variation check.

\subsubsection{\textbf{Importance of Tx Power Randomization}}
From Eq.~\eqref{eq:A-M}, we can see that if $P_{Tx}^A[i]$ was fixed and known to Supreme \textit{M}, he would be able to accurately compute the instantaneous channel fading $\chi_{AM_i}$ given that \textit{M} is not subject to any measurement error, i.e., $e_{AM_i}=0$.
\begin{equation}
    \chi_{AM_i} = P_{Tx_{i}}^{\text{A}}[i]- P_{Rx_{i}}^{\text{M}}[i] - 10 \alpha  \log_{10}\left (\frac{4\pi d_{\text{AM}_{i} }}{\lambda} \right )
\end{equation}
With the knowledge on $\chi_{AM_i}$, Supreme \textit{M} can falsify $P'^{M}_{Rx}[i]$ and $P'^{M}_{Tx}[i]$ (assuming channel reciprocity holds, i.e., $\chi_{AM_i}=\chi_{MA_i}$) as follows:
\begin{equation}
\label{e:P'^{M}_{Rx/Tx}}
    \begin{split}
      P'^{M}_{Rx}[i] = P^{M}_{Rx}[i] + 10\alpha log_{10}\left( \frac{\tilde{d}_{AM_i}}{\tilde{d}_{AB_i}}\right) + \chi_{AM_i}\\
      P'^{M}_{Tx}[i] = P^{M}_{Tx}[i] - 10\alpha log_{10}\left( \frac{\tilde{d}_{AM_i}}{\tilde{d}_{AB_i}}\right) - \chi_{MA_i}
    \end{split}
\end{equation}

Similar to the derivation of Eqs. \eqref{e:sup_PL^AM} and \eqref{e:sup_PL^MA}, we can obtain that 
\begin{equation} \label{e:sup_PL^AM/MA_fix_tx}
\begin{split}
&PL'^{AM}[i] = 10\alpha \log_{10} \left ( \frac{4 \pi d_{AB_{i}} } {\lambda} \right )\\
&PL'^{MA}[i] = 10\alpha \log_{10} \left ( \frac{4 \pi d_{AB_{i}} } {\lambda} \right ) + e_{MA_{i}}
\end{split}
\end{equation}

We can see that the computed $PL^{'AM}[i]$ is not subject to any fading or random error.
Supreme \textit{M} could add a very small randomness into the falsified report to make it look random and keep the variation below the threshold to pass the channel fading variation check.

Therefore, it is very important to randomize the Tx power at \textit{A} and \textit{B} to hide the channel fading variation from Supreme \textit{M}.
The randomization should be large enough to cover $\chi_{AM_i}$ and $\chi_{BM_i}$.
Next, we will show that Supreme \textit{M} could construct $P_{Rx}^{'M}$ and $P_{Tx}^{'M}$ in a different way from Eq.~\eqref{e:P'^{M}_{Rx/Tx}} trying to pass the channel fading variation check.
Then we will discuss what distribution the $P_{Tx}^A$ and $P_{Tx}^B$ should follow to prevent Supreme \textit{M} from passing the channel fading variation check. 

\subsubsection{\textbf{Standard Deviation of Randomized Transmission Power}}

Since the analysis on $P_{Tx}^{A}$ and $P_{Tx}^{B}$ is similar, we will use $P_{Tx}^A$ as an example.
We will first show Supreme \textit{M} can estimate the average Tx power ${\overline{P}}_{Tx}^A$ by averaging all the estimated $P_{Tx}^A[i]$ as follows:
\begin{equation} \label{eq:avg-Tx}
\begin{split}
\overline{P}_{Tx}^A 
&= \frac{1}{n} {\sum_{i=1}^{n} P_{Tx}^A[i] } = \frac{1}{n}\sum_{i=1}^{n}  \left[ P_{Rx}^M[i] + 10\alpha\log_{10} \left(\frac{4\pi d_{AM_{i}}} {\lambda}\right) + \chi_{AM_i}\right]\\
&\approx \frac{1}{n}\sum_{i=1}^{n}  \left[ P_{Rx}^M[i] + 10\alpha\log_{10} \left(\frac{4\pi d_{AM_{i}}} {\lambda}\right)\right]
\end{split},
\end{equation}
where $n$ is the total number of packets/samples.
The approximation comes from the fact that $\chi_{AM_i}$ follows zero mean Gaussian distribution, so $\frac{1}{n}\sum_{i=1}^{n} \chi_{AM_{i}}$ approaches $0$.
Therefore, Supreme~\textit{M} can falsify its Rx power as follows:
\begin{equation}
   P'^{M}_{Rx}[i] = \overline{P}_{Tx}^A  - 10\alpha\log_{10}\left ( \frac{4 \pi d_{AB_{i}}}{\lambda}\right)
\end{equation}

Then the pathloss computed at \textit{A}  will be:
\begin{equation} \label{eq:avg-Tx}
\begin{split}
PL^{'AM}[i] & = P^{A}_{Tx}[i] - P^{'M}_{Rx}[i] = P^{A}_{Tx}[i] - \overline{ P}^{A}_{Tx} + 10\alpha\log_{10}\left ( \frac{4 \pi d_{AB_{i}}}{\lambda} \right ) \\
& = 10\alpha\log_{10} 
 \left (\frac{4 \pi d_{AB_{i}}}{\lambda} \right ) + \Upsilon_{i},
\end{split}
\end{equation}
where $\Upsilon_{i}=P^{A}_{Tx}[i] - \overline{ P}^{A}_{Tx}$, which should have zero mean and follow the same distribution type as $P_{Tx}^{A}[i]$ with the same standard deviation.
Denoting the standard deviation of the random Tx power as $\sigma_{t}$, to prevent Supreme \textit{M} from passing the channel fading variation check, we should have $\sigma_{t}>\sigma_{AM}$.

Note that the threshold value in the channel fading variation check should be adapted to different environments.
In addition to distance, environmental factors can also influence channel fading. 
These environmental determinants of channel fading will be explored in detail in Section \ref{SEC:PE}.

\begin{figure*}[b] 
  \centering 
  \begin{subfigure}{.28\textwidth}
    \includegraphics[width=\linewidth]{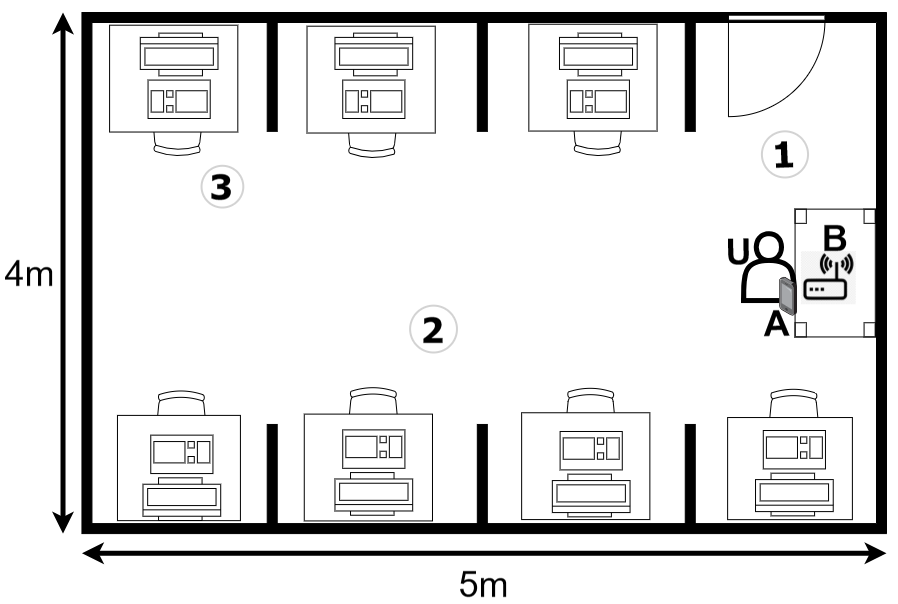}
    \caption{Confined office room}
    \label{fig:office-layout}
  \end{subfigure}%
  \hfill
  \begin{subfigure}{.34\textwidth}
    \includegraphics[width=\linewidth]{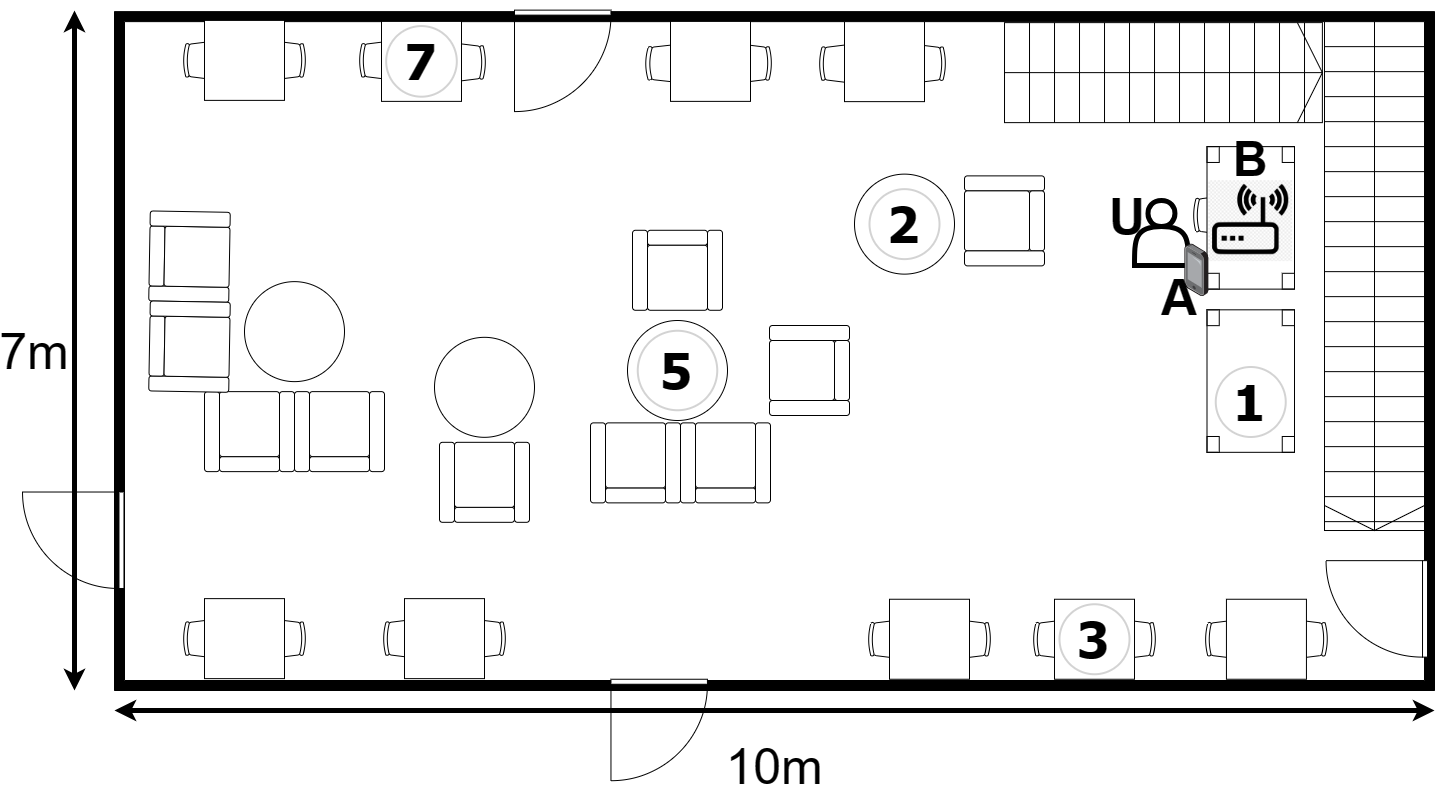}
    \caption{Spacious engineering building lobby}
    \label{fig:lobby-layout}
  \end{subfigure}%
  \hfill
  \begin{subfigure}{.34\textwidth}
    \includegraphics[width=\linewidth]{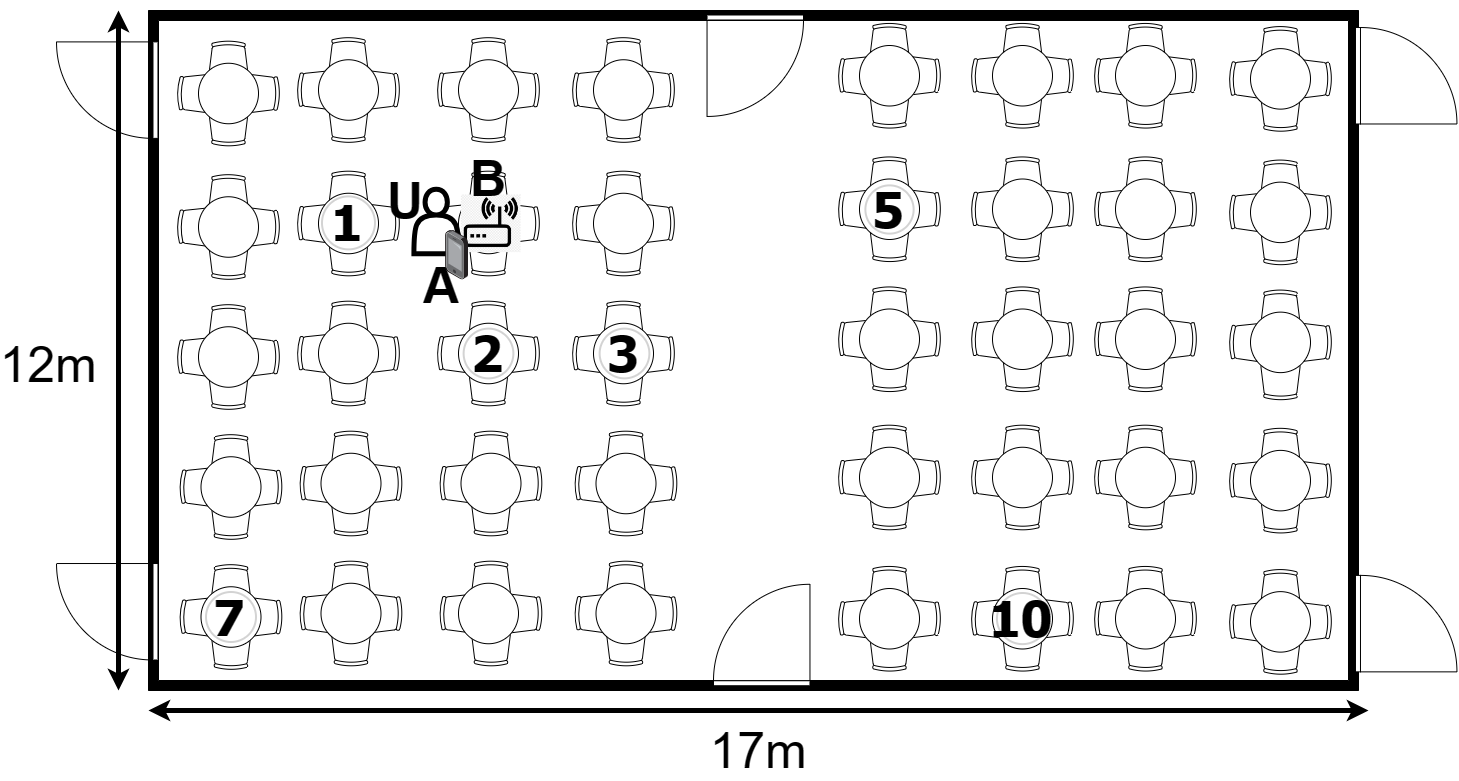}
    \caption{High-density student dining center}
    \label{fig:dining-layout}
  \end{subfigure}
  
  \begin{subfigure}{.33\textwidth}
    \includegraphics[width=\linewidth]{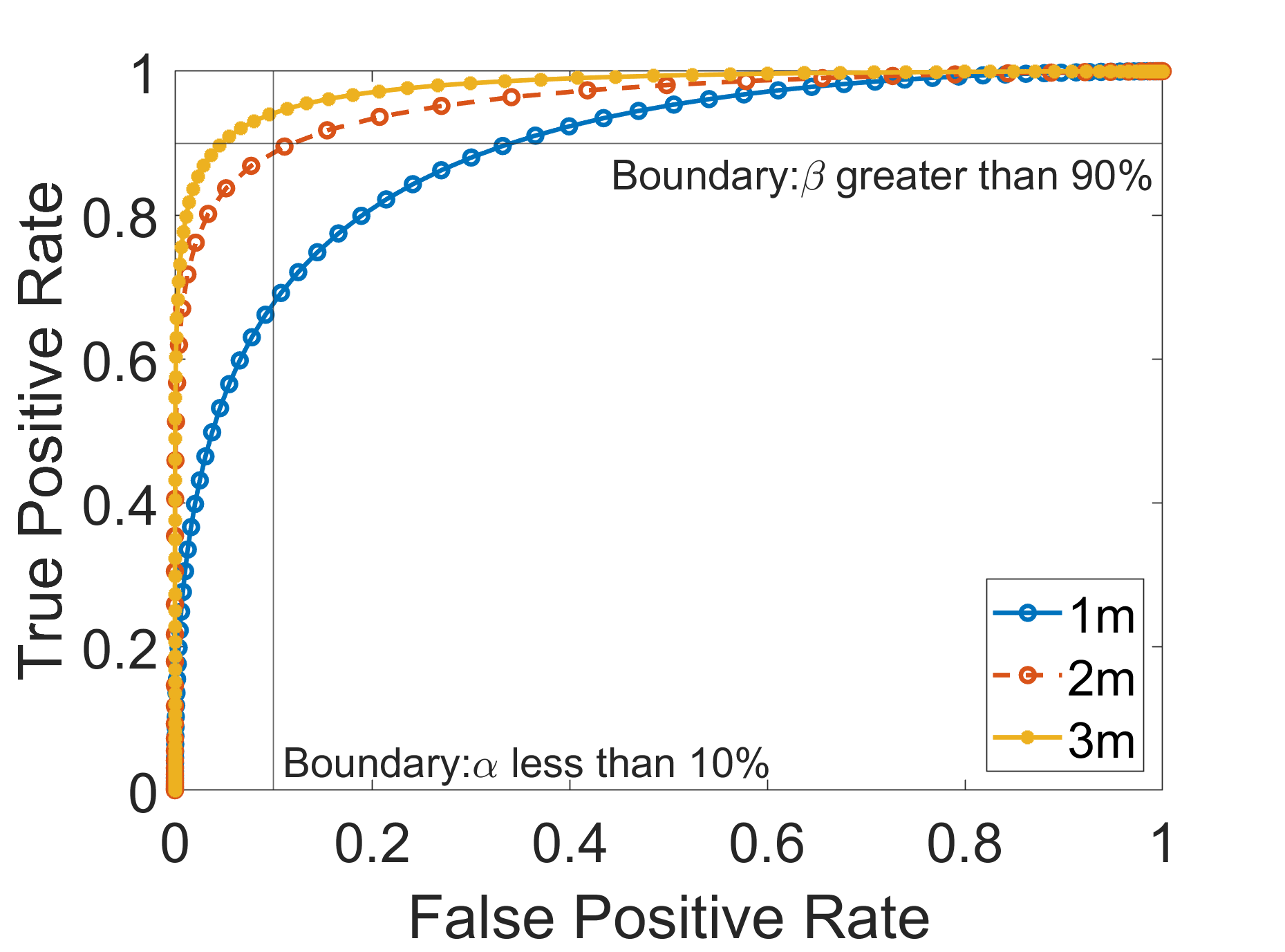}
    \caption{ROC in confined office room}
    \label{fig:Roc-office}
  \end{subfigure}%
  \hfill
  \begin{subfigure}{.33\textwidth}
    \includegraphics[width=\linewidth]{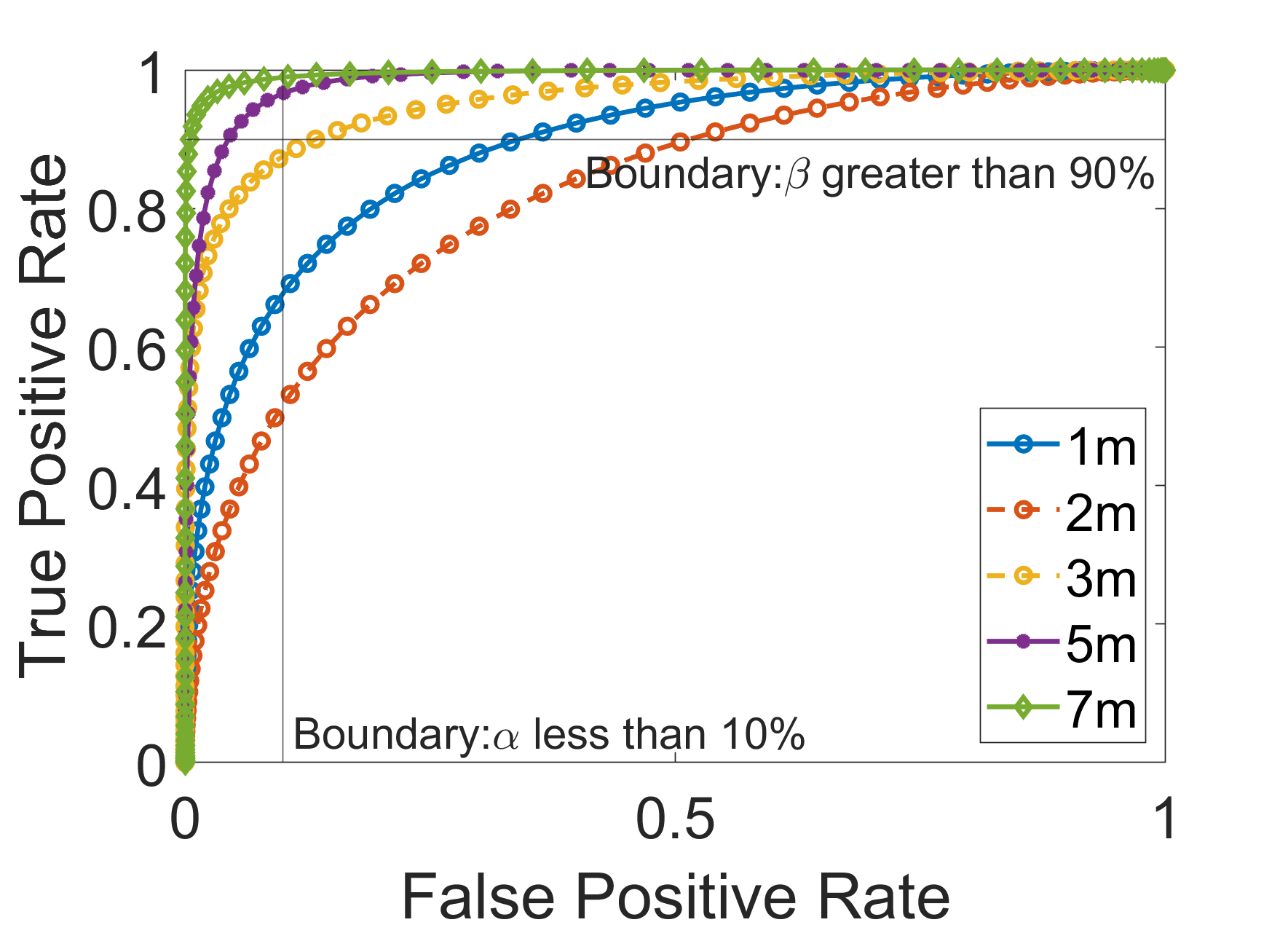}
    \caption{ROC in engineering building lobby}
    \label{fig:Roc-lobby}
  \end{subfigure}%
  \hfill
  \begin{subfigure}{.33\textwidth}
    \includegraphics[width=\linewidth]{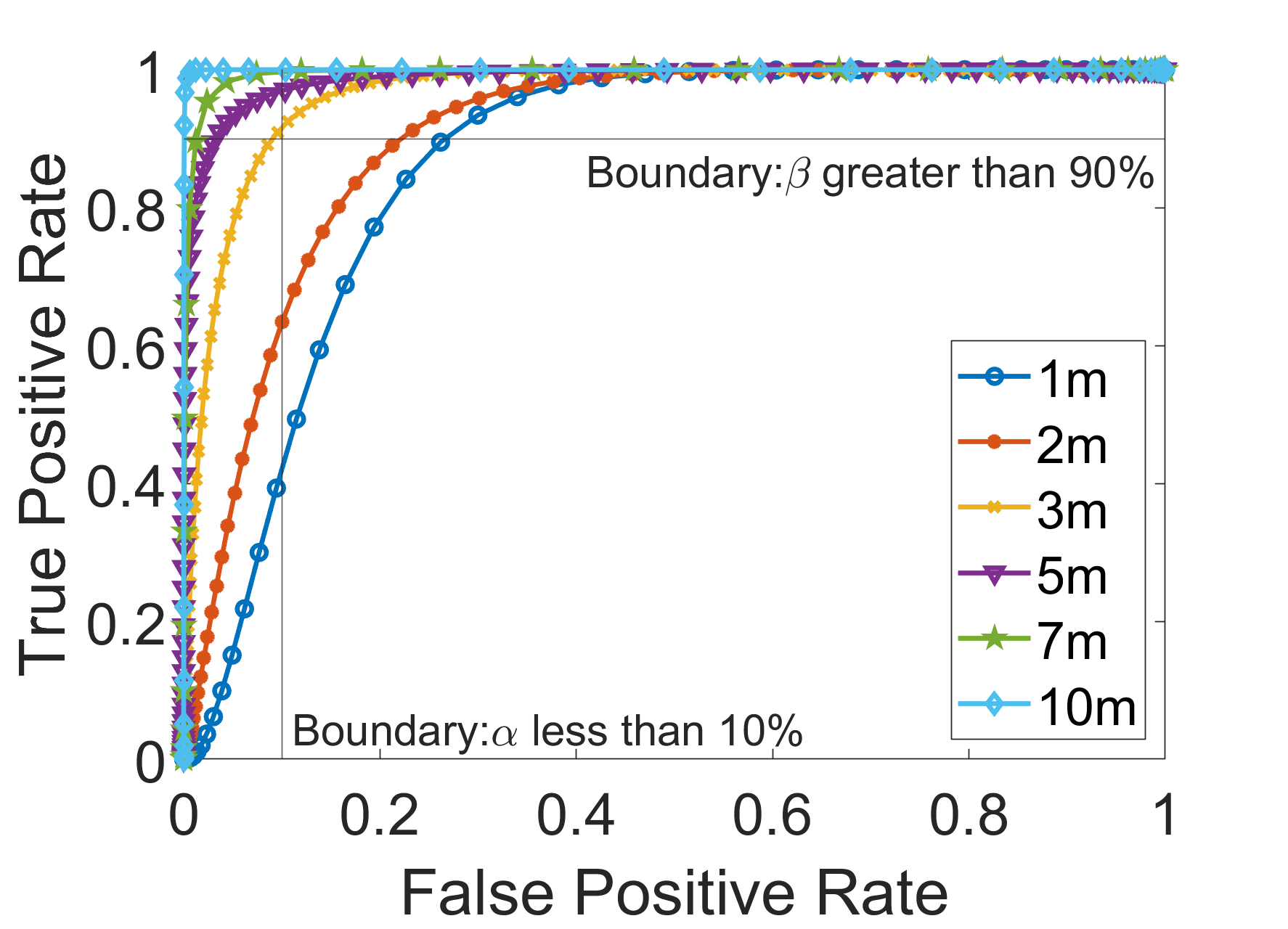}
    \caption{ROC in student dining center}
    \label{fig:Roc-HD}
  \end{subfigure}
  \caption{ROC of channel fading variation check in three different environments.}

  \label{fig:Roc}
\end{figure*}

\section{PERFORMANCE EVALUATION}
\label{SEC:PE}

In this section, we present experimental results to validate the security strength and usability of Swipe2Pair.

\vspace{-.3cm}
\subsection{Experiment Setup}

\begin{figure}[H]
    \vspace{-.4cm}
    \centering
    \includegraphics[scale=0.25]{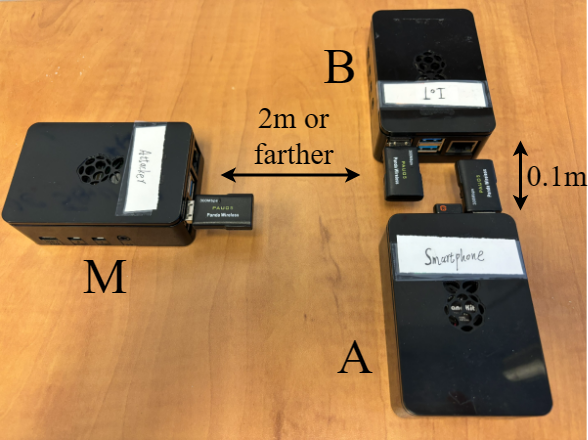}
    \caption{Layout of two adjacent pairing devices `A' and `B' and one remote attacker `M'.}
    \label{fig:devices-layout}
    \vspace{-.4cm}
\end{figure}

In our experiments, we used three Raspberry Pi4s to implement \textit{A}, \textit{B}, and \textit{M}, as illustrated in Fig.~\ref{fig:devices-layout}.
\textit{A} and \textit{B} were placed $0.1m$ from each other and \textit{M} was at least $2m$ away from them.
Each Pi4 was equipped with a PAU05 wireless dongle that was operated on channel 6 at 2.4~GHz. 
All three Pi4s were set to ad-hoc mode so they could extract the received signal strength indicator (RSSI) information for received/eavesdropped packets.
As described in Section \ref{sec:PD}, \textit{A} sent duplicated packets (including public key) at a rate of $500pkts/s$ during the swipe in front of \textit{B} from left to right.
Device~\textit{B} listened to packets from \textit{A} and then quickly sent its public key back to \textit{A} to ensure the reciprocity of the bidirectional channels largely holds. 
\textit{A} and \textit{B} randomized Tx power uniformly ranging from 0 to 30~dBm.
Since the security of Swipe2Pair is fundamentally based on the pathlosses of the legitimate channel $A-B$ and attacking channels $A-M/B-M$, it is sufficient to put \textit{M} in monitor mode to measure the RSSI for the packets sent by \textit{A/B} rather than fully implement the three types of \textit{M}.
The randomized Tx power and monitored Rx power were logged locally on each device.
The total pairing process was accomplished within one second.

\vspace{-.3cm}
\subsection{Thresholds for Security Checks} 

We illustrated the process of the Valley shape detection and check using one representative experimental data in Fig.\ref{fig:VS}. 
Initially, a valley shape was identified from noisy data in Fig.~\ref{fig:VS-detect}. The green curves in the upper subplot represent the boundary for detecting the valley shape, calculated through the first 100 data points. 
The lower subplot identifies the valley position. 
Subsequently, we use model fitting to locate the start and end points of the valley using the method proposed in \cite{o2022pragmatic} as illustrated in Fig.~\ref{fig:VS-findpoints} and applied these identified points to the original pathloss data in Fig.~\ref{fig:VS-cutpoints} to extract the range of the valley shape data in Fig.~\ref{fig:VS-rangedata}. Typically, the valley shape shows a depth of $15dBm$ with peak points at around $55dBm$ and a valley point at around $40dBm$, spanning $100 \sim 200$ data points within 0.2 seconds.
Note that no obvious valley shape was observed for the pathloss in the attacking channel.

\begin{figure}[H]
    \vspace{-.4cm}
    \centering
    \includegraphics[scale=0.185]{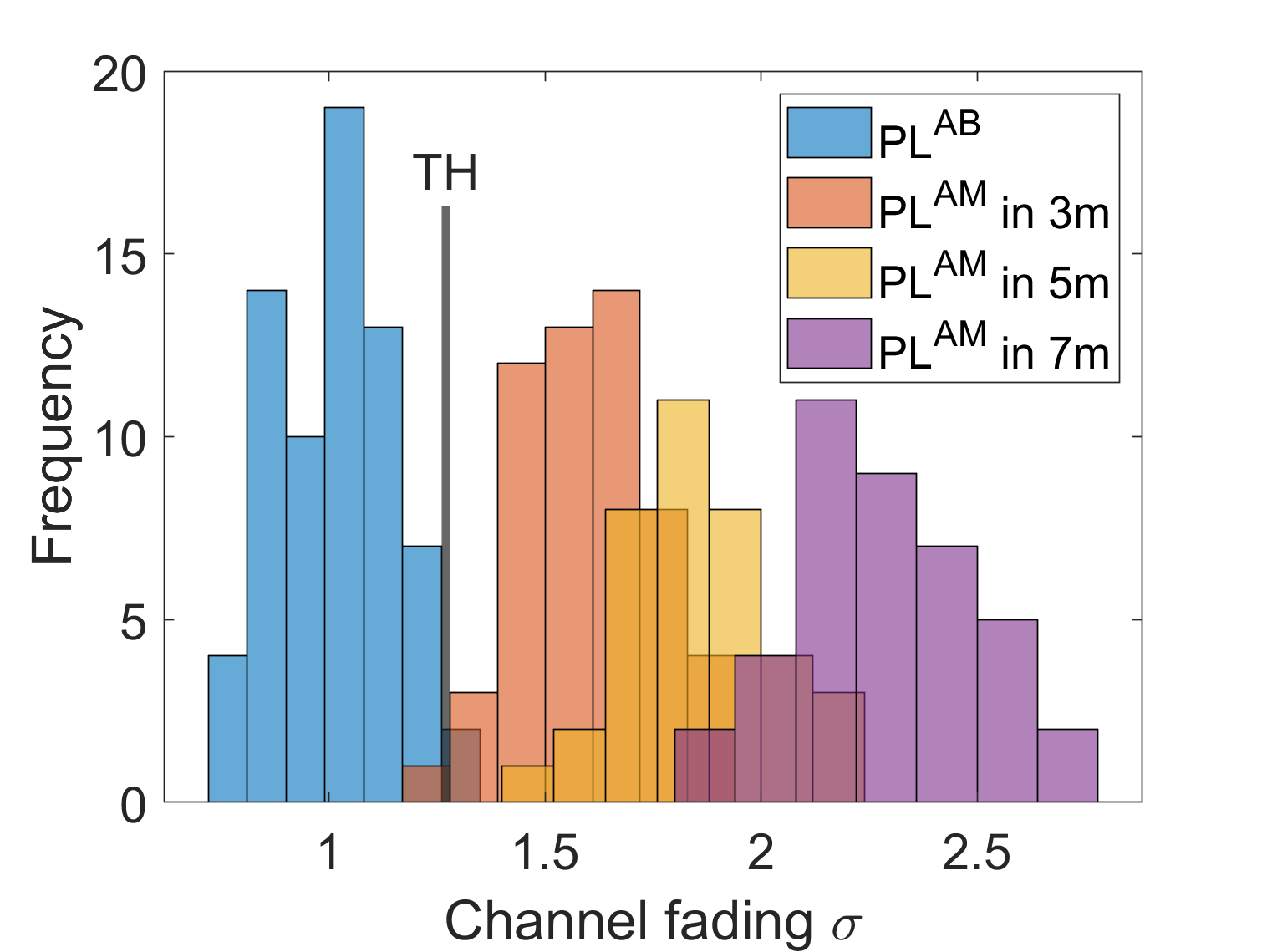}
    \vspace{-.3cm}
    \caption{Distributions of empirical standard deviations of legitimate and attacking channels at different distances.}
    \label{fig:dist-channel fading}
    \vspace{-.5cm}
\end{figure}

To validate the security of Swipe2Pair when facing a Supreme attacker, we compared the empirical standard deviations of the pathlosses of the legitimate channel $A-B$ and attack channel $A-M$ as shown in Fig.~\ref{fig:dist-channel fading}.
It can be seen that although the distance measurement error has been eliminated from supreme \textit{M}, the distributions of the empirical standard deviations of the legitimate channel and attacking channel is distinguishable. 
A threshold value of $1.27$ can effectively separate empirical standard deviation of channel $A-B$ from $A-M$. 
In addition, we tested the attacking channel across various distances in our 50 runs of experiments and found that the longer the distance from the attacker to the pairing device the larger channel variation it experiences. 
Therefore, when an attacker is further way, the attacking channel become more distinguishable from the legitimate channel in terms of channel variation. 
Besides, the environment is another factor influencing the channel fading variation $\sigma$, which will be further detailed in the subsequent discussion.

\vspace{-1cm}
\subsection{Security in Different Environments}
We evaluated the performance of channel fading variation check in three typical environments as illustrated in Fig.~\ref{fig:Roc}: the confined office room, spacious building lobby, and high-density student dining center.

\textbf{1: Confined office room.} Our office room furnishes 7 modern office table sets shown in Fig.~\ref{fig:office-layout}. 
The user (labeled as \textbf{U}) held device \textit{A} (labeled as \textbf{A}) to pair the device \textit{B} (labeled as \textbf{B}), which was stationed on a small mobile drawer measuring 0.7m in height and 0.5m in width. The device \textit{B} was centrally placed on one edge of the drawer, while \textit{A} was moved from left to right in front of  \textit{B} with a swiping distance of 0.1m at a speed of about $0.5m/s$.
The \textit{M} was placed over different positions with a similar height to \textit{A/B}, tagged in numbers, which indicate the distance in meters between  \textit{M} and the pairing devices. 
Fig.~\ref{fig:Roc-office} shows the ROC (receiver operating characteristic) curve of detecting the attacker. 
It can be seen that the detection performance becomes better when the attacker is farther away from the pairing devices. 
To achieve a desirable false positive rate (e.g., $\alpha<10\%$) and true positive rate (e.g., $\beta>90\%$), a separation of $2m$ between the attacker and the pairing devices is sufficient in the tested confined office room.

\textbf{2: Spacious engineering building lobby.} Our engineering building lobby represents a more spacious environment filled with multiple 3-piece round bar table sets and coffee tables in Fig. \ref{fig:lobby-layout}. In this setting, devices \textit{A} and \textit{B} were placed in proximity over a table with a distance of $0.1m$, while \textit{M} was located at different distances from the pairing devices.
Compared to the confined office, similar detection performance improvement over longer distances is observed in Fig.~\ref{fig:Roc-lobby}, while a distance of at least $3m$ between the attacker and pairing devices is needed to achieve a desirable security strength (i.e., $\alpha<10\%$ and $\beta>90\%$). 
A spacious lobby diminishes reflections from walls, yet more obstacles and human activities enhance the shadowing effect in a longer distance.

\textbf{3: High-density student dining center.} The last environment is the crowded student dining center with many round dinner tables surrounded by four chairs each in Fig.~\ref{fig:dining-layout}. 
We placed devices \textit{A} and \textit{B} on one table with similar operation to previous experiments, and placed \textit{M} on different tables at different locations in the dining hall. 
Similarly, we observed performance improvement along a longer distance between \textit{M} and \textit{A/B} as shown in Fig.~\ref{fig:Roc-HD}.
It is again verified that the channel fading variation is increased as \textit{M} was placed progressively farther away from \textit{A/B}. 
A distance of at least $3m$ is also sufficient to achieve a desirable security strength. 

Moreover, we observed that the channel fading variation is intensified significantly beyond a certain distance (e.g., $3m$) in both the lobby and dining center. 
We attribute these observations to the complex environment caused by furniture (e.g., tables and chairs) and human activities that result in a rich multi-path radio environment \cite{hamida2010investigating}. 
The pronounced channel fading variation leads to improved attack detection performance in the channel fading variation check.

\vspace{-.3cm}
\subsection{Robustness of Valley Shape Detection}

\begin{figure}
    \vspace{-.5cm}
    \centering
    \includegraphics[scale=0.20]{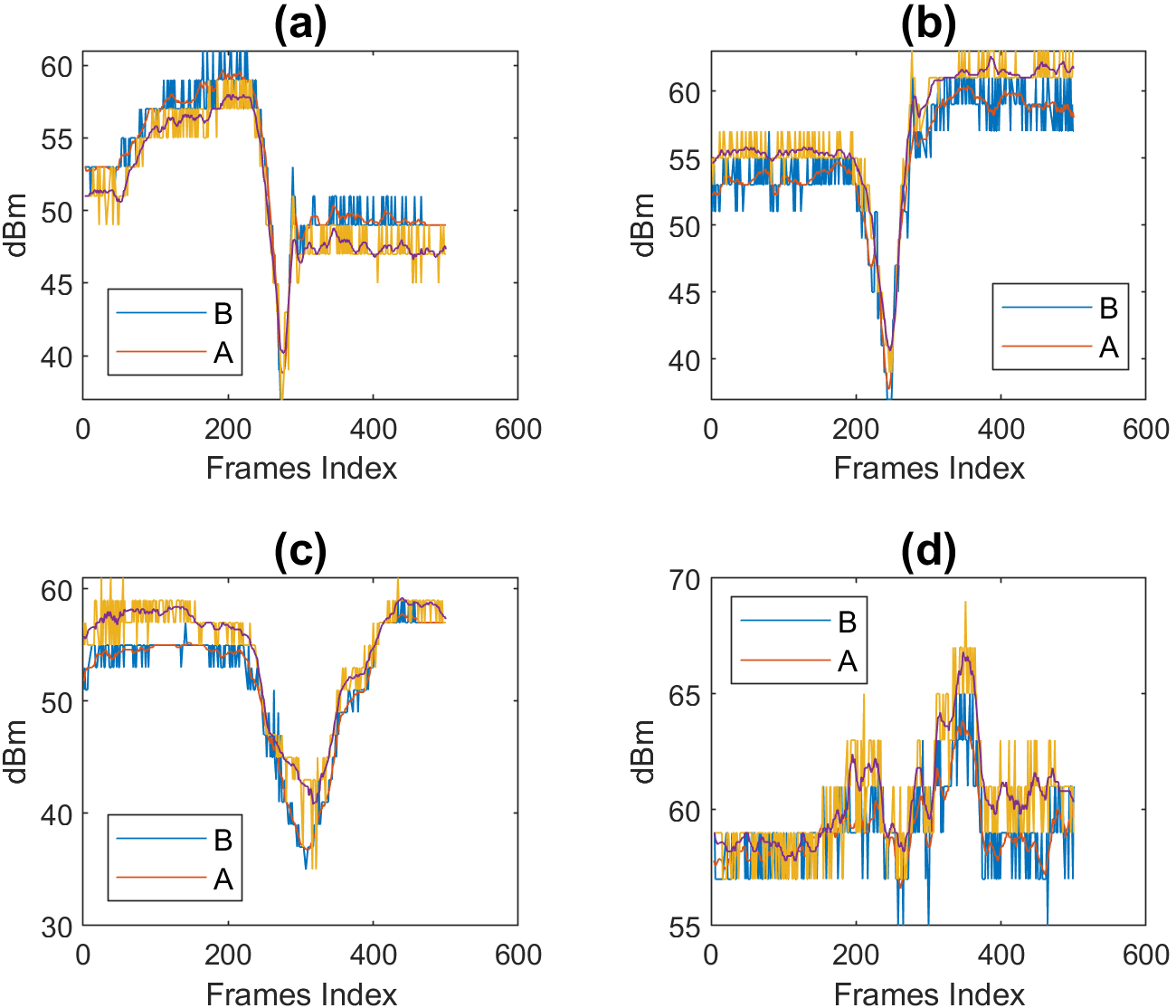}
    \vspace{-.3cm}
    \caption{Pathloss of the channel between adjacent pairing devices for different imperfect swiping}
    \label{fig:4vs}
    \vspace{-.6cm}
\end{figure}

We investigated the robustness of the valley shape detection against imperfect human motions.
We examined valley shapes generated from four different imperfect pairing motions.
The two pairing devices were placed within $10cm$ of each other with a $500pkts/s$ packet sending rate over a 1-second pairing duration.

We observed that a valley shape could be detected through  Algorithm~\ref{alg:valley-detection} under different imperfect motions caused by human.
In Fig.~\ref{fig:4vs}: (a) was conducted through asymmetric swiping where device \textit{B} was not at a center point when \textit{A} is swiping in front of it; (b) was the diagonal swiping where \textit{A} is swiping from the right bottom to the left top of device \textit{B}; (c) was the slow swiping where \textit{A} swipes slowly in front of \textit{B}; and (d) was a swiping not in close proximity where the distance between the two pairing devices is larger than $50cm$. 

From Fig.~\ref{fig:4vs}, we can see that the valley shape is detectable across various imperfect human motions whether swiping with device misalignment (Fig.~\ref{fig:4vs}(a)), or in different directions (Fig.~\ref{fig:4vs}(b)) or speeds (Fig.~\ref{fig:4vs}(c)). 
However, we observed that the valley shape in (d) became flatter and less likely to be recognized.
This observation validates that the valley shape check is a good test to differentiate a close-by pairing device from an attacker that is not in close proximity.

\vspace{-.2cm}
\section{Conclusion}

We introduced Swipe2Pair as a novel and fast approach for achieving secure in-band pairing of wireless devices without extra sensors or sophisticated user input/output interfaces. 
Swipe2Pair provides a ubiquitous solution for in-band close-by wireless device pairing as it only relies on wireless interfaces and a simple user motion, i.e., swiping one pairing device in front of the other within $1s$.
Strong security is achieved by utilizing random transmission power to hide the channel fading information from the attacker and leveraging the property of wireless fading channel to differentiate pairing devices in close proximity from attackers who are at a distance from the pairing devices at least $2-3m$ away.

We conducted a comprehensive security analysis of Swipe2Pair considering three levels of attacker capability: General, Advanced, and Supreme.  
The security
performance of Swipe2Pair was further validated through experiments in different environments with various settings and under imperfect human motions. 
Our analysis and experimental results
demonstrate the robustness, reliability, and efficiency of Swipe2Pair
as a secure in-band device pairing protocol.
Even a Supreme attacker who can accurately estimate the distance and motion of pairing devices is unable to compromise Swipe2Pair.

\vspace{-.2cm}
\section{ACKNOWLEDGEMENT}
This material is based on research sponsored by the Defense Advanced Research Projects Agency (DARPA) under the agreement number HR001120C0154. 
The views, opinions, and/or findings contained in this article are those of the author(s) and should not be interpreted as representing the official views or policies, either expressed or implied, of the Defense Advanced Research Projects Agency or the Department of Defense.

\clearpage

\balance
\bibliographystyle{ACM-Reference-Format}
\bibliography{IEEEabrv,S2P}


\begin{thebibliography}{38}


\ifx \showCODEN    \undefined \def \showCODEN     #1{\unskip}     \fi
\ifx \showDOI      \undefined \def \showDOI       #1{#1}\fi
\ifx \showISBNx    \undefined \def \showISBNx     #1{\unskip}     \fi
\ifx \showISBNxiii \undefined \def \showISBNxiii  #1{\unskip}     \fi
\ifx \showISSN     \undefined \def \showISSN      #1{\unskip}     \fi
\ifx \showLCCN     \undefined \def \showLCCN      #1{\unskip}     \fi
\ifx \shownote     \undefined \def \shownote      #1{#1}          \fi
\ifx \showarticletitle \undefined \def \showarticletitle #1{#1}   \fi
\ifx \showURL      \undefined \def \showURL       {\relax}        \fi
\providecommand\bibfield[2]{#2}
\providecommand\bibinfo[2]{#2}
\providecommand\natexlab[1]{#1}
\providecommand\showeprint[2][]{arXiv:#2}

\bibitem[Adewumi et~al\mbox{.}(2013)]%
        {adewumi2013rssi}
\bibfield{author}{\bibinfo{person}{Omotayo~G Adewumi}, \bibinfo{person}{Karim Djouani}, {and} \bibinfo{person}{Anish~M Kurien}.} \bibinfo{year}{2013}\natexlab{}.
\newblock \showarticletitle{RSSI based indoor and outdoor distance estimation for localization in WSN}. In \bibinfo{booktitle}{\emph{2013 IEEE International Conference on Industrial Technology (ICIT)}}. IEEE, \bibinfo{pages}{1534--1539}.
\newblock


\bibitem[Bellovin and Merritt(1994)]%
        {bellovin1994attack}
\bibfield{author}{\bibinfo{person}{Steven~M Bellovin} {and} \bibinfo{person}{Michael Merritt}.} \bibinfo{year}{1994}\natexlab{}.
\newblock \showarticletitle{An attack on the interlock protocol when used for authentication}.
\newblock \bibinfo{journal}{\emph{IEEE Transactions on Information Theory}} \bibinfo{volume}{40}, \bibinfo{number}{1} (\bibinfo{year}{1994}), \bibinfo{pages}{273--275}.
\newblock


\bibitem[Brakel(2014)]%
        {brakel2014}
\bibfield{author}{\bibinfo{person}{J.P.G~van Brakel}.} \bibinfo{year}{2014}\natexlab{}.
\newblock \bibinfo{title}{Robust peak detection algorithm using z-scores}.
\newblock \bibinfo{howpublished}{https://stackoverflow.com/questions/22583391/peak-signal-detection-in-realtime-timeseries-data/22640362\#22640362}.
\newblock
\urldef\tempurl%
\url{https://stackoverflow.com/questions/22583391/peak-signal-detection-in-realtime-timeseries-data/22640362\#22640362}
\showURL{%
\tempurl}


\bibitem[Cai et~al\mbox{.}(2011)]%
        {Cai:2011}
\bibfield{author}{\bibinfo{person}{Liang Cai}, \bibinfo{person}{Kai Zeng}, \bibinfo{person}{Hao Chen}, {and} \bibinfo{person}{Prasant Mohapatra}.} \bibinfo{year}{2011}\natexlab{}.
\newblock \showarticletitle{{Good Neighbor: Secure pairing of nearby wireless devices by multiple antennas}}. In \bibinfo{booktitle}{\emph{NDSS}}.
\newblock


\bibitem[Gallo et~al\mbox{.}(2008)]%
        {4563424}
\bibfield{author}{\bibinfo{person}{Michele Gallo}, \bibinfo{person}{Peter~S. Hall}, \bibinfo{person}{Yuriy~I. Nechayev}, {and} \bibinfo{person}{Michele Bozzetti}.} \bibinfo{year}{2008}\natexlab{}.
\newblock \showarticletitle{Use of Animation Software in Simulation of On-Body Communications Channels at 2.45 GHz}.
\newblock \bibinfo{journal}{\emph{IEEE Antennas and Wireless Propagation Letters}}  \bibinfo{volume}{7} (\bibinfo{year}{2008}), \bibinfo{pages}{321--324}.
\newblock
\urldef\tempurl%
\url{https://doi.org/10.1109/LAWP.2008.928484}
\showDOI{\tempurl}


\bibitem[Ghose et~al\mbox{.}(2017)]%
        {ghose2017help}
\bibfield{author}{\bibinfo{person}{Nirnimesh Ghose}, \bibinfo{person}{Loukas Lazos}, {and} \bibinfo{person}{Ming Li}.} \bibinfo{year}{2017}\natexlab{}.
\newblock \showarticletitle{$\{$HELP$\}$:$\{$Helper-Enabled$\}$$\{$In-Band$\}$ Device Pairing Resistant Against Signal Cancellation}. In \bibinfo{booktitle}{\emph{26th USENIX Security Symposium (USENIX Security 17)}}. \bibinfo{pages}{433--450}.
\newblock


\bibitem[Ghose et~al\mbox{.}(2020)]%
        {ghose2020band}
\bibfield{author}{\bibinfo{person}{Nirnimesh Ghose}, \bibinfo{person}{Loukas Lazos}, {and} \bibinfo{person}{Ming Li}.} \bibinfo{year}{2020}\natexlab{}.
\newblock \showarticletitle{In-band secret-free pairing for COTS wireless devices}.
\newblock \bibinfo{journal}{\emph{IEEE Transactions on Mobile Computing}} \bibinfo{volume}{21}, \bibinfo{number}{2} (\bibinfo{year}{2020}), \bibinfo{pages}{612--628}.
\newblock


\bibitem[Gong and Haenggi(2014)]%
        {6380497}
\bibfield{author}{\bibinfo{person}{Zhenhua Gong} {and} \bibinfo{person}{Martin Haenggi}.} \bibinfo{year}{2014}\natexlab{}.
\newblock \showarticletitle{Interference and Outage in Mobile Random Networks: Expectation, Distribution, and Correlation}.
\newblock \bibinfo{journal}{\emph{IEEE Transactions on Mobile Computing}} \bibinfo{volume}{13}, \bibinfo{number}{2} (\bibinfo{year}{2014}), \bibinfo{pages}{337--349}.
\newblock
\urldef\tempurl%
\url{https://doi.org/10.1109/TMC.2012.253}
\showDOI{\tempurl}


\bibitem[Hamida and Chelius(2010)]%
        {hamida2010investigating}
\bibfield{author}{\bibinfo{person}{Elyes~Ben Hamida} {and} \bibinfo{person}{Guillaume Chelius}.} \bibinfo{year}{2010}\natexlab{}.
\newblock \showarticletitle{Investigating the impact of human activity on the performance of wireless networks - An experimental approach}. In \bibinfo{booktitle}{\emph{2010 IEEE International Symposium on A World of Wireless, Mobile and Multimedia Networks (WoWMoM)}}. IEEE, \bibinfo{pages}{1--8}.
\newblock


\bibitem[Han et~al\mbox{.}(2018)]%
        {Han:2018}
\bibfield{author}{\bibinfo{person}{Jun Han}, \bibinfo{person}{Albert~Jin Chung}, \bibinfo{person}{Manal~Kumar Sinha}, \bibinfo{person}{Madhumitha Harishankar}, \bibinfo{person}{Shijia Pan}, \bibinfo{person}{Hae~Young Noh}, \bibinfo{person}{Pei Zhang}, {and} \bibinfo{person}{Patrick Tague}.} \bibinfo{year}{2018}\natexlab{}.
\newblock \showarticletitle{{Do you feel what I hear? Enabling autonomous IoT device pairing using different sensor types}}. In \bibinfo{booktitle}{\emph{IEEE Symposium on Security and Privacy (SP)}}. \bibinfo{pages}{836--852}.
\newblock


\bibitem[Ibrahim et~al\mbox{.}(2022)]%
        {ibrahim2022eolo}
\bibfield{author}{\bibinfo{person}{Omar~Adel Ibrahim}, \bibinfo{person}{Gabriele Oligeri}, {and} \bibinfo{person}{Roberto Di~Pietro}.} \bibinfo{year}{2022}\natexlab{}.
\newblock \showarticletitle{Eolo: IoT Proximity-based Authentication via Pressure Correlated Variations}. In \bibinfo{booktitle}{\emph{2022 IEEE Conference on Communications and Network Security (CNS)}}. IEEE, \bibinfo{pages}{109--117}.
\newblock


\bibitem[Jin et~al\mbox{.}(2014)]%
        {jin2014magpairing}
\bibfield{author}{\bibinfo{person}{Rong Jin}, \bibinfo{person}{Liu Shi}, \bibinfo{person}{Kai Zeng}, \bibinfo{person}{Amit Pande}, {and} \bibinfo{person}{Prasant Mohapatra}.} \bibinfo{year}{2014}\natexlab{}.
\newblock \showarticletitle{MagPairing: Exploiting magnetometers for pairing smartphones in close proximity}. In \bibinfo{booktitle}{\emph{IEEE Conf. on Comm. and Netw. Security}}. \bibinfo{pages}{445--453}.
\newblock


\bibitem[Khalfaoui et~al\mbox{.}(2021)]%
        {khalfaoui2021security}
\bibfield{author}{\bibinfo{person}{Sameh Khalfaoui}, \bibinfo{person}{Jean Leneutre}, \bibinfo{person}{Arthur Villard}, \bibinfo{person}{Jingxuan Ma}, {and} \bibinfo{person}{Pascal Urien}.} \bibinfo{year}{2021}\natexlab{}.
\newblock \showarticletitle{Security analysis of out-of-band device pairing protocols: a survey}.
\newblock \bibinfo{journal}{\emph{Wireless Communications and Mobile Computing}}  \bibinfo{volume}{2021} (\bibinfo{year}{2021}), \bibinfo{pages}{1--30}.
\newblock


\bibitem[Kumar and Sukumar(2019)]%
        {kumar2019achieving}
\bibfield{author}{\bibinfo{person}{K.~S. Kumar} {and} \bibinfo{person}{R. Sukumar}.} \bibinfo{year}{2019}\natexlab{}.
\newblock \showarticletitle{{Achieving energy efficiency using novel scalar multiplication based ECC for Android devices in Internet of Things environments}}.
\newblock \bibinfo{journal}{\emph{Cluster Computing}} \bibinfo{volume}{22}, \bibinfo{number}{5} (\bibinfo{year}{2019}), \bibinfo{pages}{12021--12028}.
\newblock


\bibitem[Li et~al\mbox{.}(2020)]%
        {Li:2020}
\bibfield{author}{\bibinfo{person}{Xiaopeng Li}, \bibinfo{person}{Qiang Zeng}, \bibinfo{person}{Lannan Luo}, {and} \bibinfo{person}{Tongbo Luo}.} \bibinfo{year}{2020}\natexlab{}.
\newblock \showarticletitle{{T2pair: Secure and usable pairing for heterogeneous IoT devices}}. In \bibinfo{booktitle}{\emph{ACM SIGSAC}}. \bibinfo{pages}{309--323}.
\newblock


\bibitem[Mayrhofer and Gellersen(2009)]%
        {Mayrhofer:2009}
\bibfield{author}{\bibinfo{person}{Rene Mayrhofer} {and} \bibinfo{person}{Hans Gellersen}.} \bibinfo{year}{2009}\natexlab{}.
\newblock \showarticletitle{{Shake well before use: Intuitive and secure pairing of mobile devices}}.
\newblock \bibinfo{journal}{\emph{{IEEE} Trans. Mobile Comput.}} \bibinfo{volume}{8}, \bibinfo{number}{6} (\bibinfo{year}{2009}), \bibinfo{pages}{792--806}.
\newblock


\bibitem[McGinthy and Michaels(2018)]%
        {mcginthy2018session}
\bibfield{author}{\bibinfo{person}{Jason McGinthy} {and} \bibinfo{person}{Alan Michaels}.} \bibinfo{year}{2018}\natexlab{}.
\newblock \showarticletitle{Session key derivation for low power IoT devices}. In \bibinfo{booktitle}{\emph{2018 IEEE 4th International Conference on Big Data Security on Cloud (BigDataSecurity), IEEE International Conference on High Performance and Smart Computing,(HPSC) and IEEE International Conference on Intelligent Data and Security (IDS)}}. IEEE, \bibinfo{pages}{194--203}.
\newblock


\bibitem[Mei et~al\mbox{.}(2019)]%
        {mei2019listen}
\bibfield{author}{\bibinfo{person}{Shijia Mei}, \bibinfo{person}{Zhihong Liu}, \bibinfo{person}{Yong Zeng}, \bibinfo{person}{Lin Yang}, {and} \bibinfo{person}{Jian~Feng Ma}.} \bibinfo{year}{2019}\natexlab{}.
\newblock \showarticletitle{Listen!: Audio-based smart iot device pairing protocol}. In \bibinfo{booktitle}{\emph{2019 IEEE 19th International Conference on Communication Technology (ICCT)}}. IEEE, \bibinfo{pages}{391--397}.
\newblock


\bibitem[Mohammed(2016)]%
        {mohammed2016distance}
\bibfield{author}{\bibinfo{person}{Salim~Latif Mohammed}.} \bibinfo{year}{2016}\natexlab{}.
\newblock \showarticletitle{Distance estimation based on RSSI and Log-Normal shadowing models for ZigBee wireless sensor network}.
\newblock \bibinfo{journal}{\emph{Engineering and Technology Journal}} \bibinfo{volume}{34}, \bibinfo{number}{15} (\bibinfo{year}{2016}), \bibinfo{pages}{2950--2959}.
\newblock


\bibitem[O'Haver(2022)]%
        {o2022pragmatic}
\bibfield{author}{\bibinfo{person}{T O'Haver}.} \bibinfo{year}{2022}\natexlab{}.
\newblock \bibinfo{title}{Pragmatic Introduction to Signal Processing: Applications in scientific measurement: An illustrated handbook with free software and spreadsheet templates to download}.
\newblock
\newblock


\bibitem[Pourbemany et~al\mbox{.}(2022)]%
        {pourbemany2022breathe}
\bibfield{author}{\bibinfo{person}{Jafar Pourbemany}, \bibinfo{person}{Ye Zhu}, {and} \bibinfo{person}{Riccardo Bettati}.} \bibinfo{year}{2022}\natexlab{}.
\newblock \showarticletitle{Breathe-to-Pair (B2P) Respiration-Based Pairing Protocol for Wearable Devices}. In \bibinfo{booktitle}{\emph{Proceedings of the 15th ACM Conference on Security and Privacy in Wireless and Mobile Networks}}. \bibinfo{pages}{188--200}.
\newblock


\bibitem[Rivest and Shamir(1984)]%
        {rivest1984expose}
\bibfield{author}{\bibinfo{person}{Ronald~L Rivest} {and} \bibinfo{person}{Adi Shamir}.} \bibinfo{year}{1984}\natexlab{}.
\newblock \showarticletitle{How to expose an eavesdropper}.
\newblock \bibinfo{journal}{\emph{Commun. ACM}} \bibinfo{volume}{27}, \bibinfo{number}{4} (\bibinfo{year}{1984}), \bibinfo{pages}{393--394}.
\newblock


\bibitem[Roeschlin et~al\mbox{.}(2018)]%
        {Roeschlin:2018}
\bibfield{author}{\bibinfo{person}{Marc Roeschlin}, \bibinfo{person}{Ivan Martinovic}, {and} \bibinfo{person}{Kasper~Bonne Rasmussen}.} \bibinfo{year}{2018}\natexlab{}.
\newblock \showarticletitle{Device Pairing at the Touch of an Electrode}. In \bibinfo{booktitle}{\emph{NDSS}}, Vol.~\bibinfo{volume}{18}. \bibinfo{pages}{18--21}.
\newblock


\bibitem[Rostami et~al\mbox{.}(2013)]%
        {rostami2013heart}
\bibfield{author}{\bibinfo{person}{Masoud Rostami}, \bibinfo{person}{Ari Juels}, {and} \bibinfo{person}{Farinaz Koushanfar}.} \bibinfo{year}{2013}\natexlab{}.
\newblock \showarticletitle{Heart-to-heart (H2H) authentication for implanted medical devices}. In \bibinfo{booktitle}{\emph{Proceedings of the 2013 ACM SIGSAC conference on Computer \& communications security}}. \bibinfo{pages}{1099--1112}.
\newblock


\bibitem[Saloni and Hegde(2016)]%
        {saloni2016wifi}
\bibfield{author}{\bibinfo{person}{Shubham Saloni} {and} \bibinfo{person}{Achyut Hegde}.} \bibinfo{year}{2016}\natexlab{}.
\newblock \showarticletitle{{WiFi-aware as a connectivity solution for IoT pairing IoT with WiFi aware technology: Enabling new proximity based services}}. In \bibinfo{booktitle}{\emph{IEEE IOTA}}. \bibinfo{pages}{137--142}.
\newblock


\bibitem[Shang et~al\mbox{.}(2014)]%
        {shang2014location}
\bibfield{author}{\bibinfo{person}{Fengjun Shang}, \bibinfo{person}{Wen Su}, \bibinfo{person}{Qian Wang}, \bibinfo{person}{Hongxia Gao}, {and} \bibinfo{person}{Qiang Fu}.} \bibinfo{year}{2014}\natexlab{}.
\newblock \showarticletitle{A location estimation algorithm based on RSSI vector similarity degree}.
\newblock \bibinfo{journal}{\emph{International Journal of Distributed Sensor Networks}} \bibinfo{volume}{10}, \bibinfo{number}{8} (\bibinfo{year}{2014}), \bibinfo{pages}{371350}.
\newblock


\bibitem[Siddiqi et~al\mbox{.}(2021)]%
        {siddiqi2021securing}
\bibfield{author}{\bibinfo{person}{Muhammad~Ali Siddiqi}, \bibinfo{person}{Robert~HSH Beurskens}, \bibinfo{person}{Pieter Kruizinga}, \bibinfo{person}{Chris~I De~Zeeuw}, {and} \bibinfo{person}{Christos Strydis}.} \bibinfo{year}{2021}\natexlab{}.
\newblock \showarticletitle{Securing implantable medical devices using ultrasound waves}.
\newblock \bibinfo{journal}{\emph{IEEE Access}}  \bibinfo{volume}{9} (\bibinfo{year}{2021}), \bibinfo{pages}{80170--80182}.
\newblock


\bibitem[Vahdati et~al\mbox{.}(2019)]%
        {vahdati2019comparison}
\bibfield{author}{\bibinfo{person}{Zeinab Vahdati}, \bibinfo{person}{Sharifah Yasin}, \bibinfo{person}{Ali Ghasempour}, {and} \bibinfo{person}{Mohammad Salehi}.} \bibinfo{year}{2019}\natexlab{}.
\newblock \showarticletitle{Comparison of ECC and RSA algorithms in IoT devices}.
\newblock \bibinfo{journal}{\emph{Journal of Theoretical and Applied Information Technology}} \bibinfo{volume}{97}, \bibinfo{number}{16} (\bibinfo{year}{2019}).
\newblock


\bibitem[Wang et~al\mbox{.}(2015)]%
        {wang2015wave}
\bibfield{author}{\bibinfo{person}{Wei Wang}, \bibinfo{person}{Zhan Wang}, \bibinfo{person}{Wen~Tao Zhu}, {and} \bibinfo{person}{Lei Wang}.} \bibinfo{year}{2015}\natexlab{}.
\newblock \showarticletitle{WAVE: Secure wireless pairing exploiting human body movements}. In \bibinfo{booktitle}{\emph{2015 IEEE Trustcom/BigDataSE/ISPA}}, Vol.~\bibinfo{volume}{1}. IEEE, \bibinfo{pages}{1243--1248}.
\newblock


\bibitem[Wang et~al\mbox{.}(2012)]%
        {wang2012empirical}
\bibfield{author}{\bibinfo{person}{Ye Wang}, \bibinfo{person}{Wen-jun Lu}, \bibinfo{person}{Hong-bo Zhu}, {et~al\mbox{.}}} \bibinfo{year}{2012}\natexlab{}.
\newblock \showarticletitle{An empirical path-loss model for wireless channels in indoor short-range office environment}.
\newblock \bibinfo{journal}{\emph{International Journal of Antennas and Propagation}}  \bibinfo{volume}{2012} (\bibinfo{year}{2012}).
\newblock


\bibitem[Wu et~al\mbox{.}(2017)]%
        {wu2017attack}
\bibfield{author}{\bibinfo{person}{Yongdong Wu}, \bibinfo{person}{Binbin Chen}, \bibinfo{person}{Zhigang Zhao}, {and} \bibinfo{person}{Yao Cheng}.} \bibinfo{year}{2017}\natexlab{}.
\newblock \showarticletitle{Attack and countermeasure on interlock-based device pairing schemes}.
\newblock \bibinfo{journal}{\emph{IEEE Transactions on Information Forensics and Security}} \bibinfo{volume}{13}, \bibinfo{number}{3} (\bibinfo{year}{2017}), \bibinfo{pages}{745--757}.
\newblock


\bibitem[Xu et~al\mbox{.}(2017)]%
        {xu2017gait}
\bibfield{author}{\bibinfo{person}{Weitao Xu}, \bibinfo{person}{Chitra Javali}, \bibinfo{person}{Girish Revadigar}, \bibinfo{person}{Chengwen Luo}, \bibinfo{person}{Neil Bergmann}, {and} \bibinfo{person}{Wen Hu}.} \bibinfo{year}{2017}\natexlab{}.
\newblock \showarticletitle{Gait-key: A gait-based shared secret key generation protocol for wearable devices}.
\newblock \bibinfo{journal}{\emph{ACM Transactions on Sensor Networks (TOSN)}} \bibinfo{volume}{13}, \bibinfo{number}{1} (\bibinfo{year}{2017}), \bibinfo{pages}{1--27}.
\newblock


\bibitem[Yan et~al\mbox{.}(2019)]%
        {yan2019towards}
\bibfield{author}{\bibinfo{person}{Zhenyu Yan}, \bibinfo{person}{Qun Song}, \bibinfo{person}{Rui Tan}, \bibinfo{person}{Yang Li}, {and} \bibinfo{person}{Adams Wai~Kin Kong}.} \bibinfo{year}{2019}\natexlab{}.
\newblock \showarticletitle{Towards touch-to-access device authentication using induced body electric potentials}. In \bibinfo{booktitle}{\emph{The 25th Annual International Conference on Mobile Computing and Networking}}. \bibinfo{pages}{1--16}.
\newblock


\bibitem[Yang et~al\mbox{.}(2017)]%
        {Yang2017IoTJ}
\bibfield{author}{\bibinfo{person}{Yuchen Yang}, \bibinfo{person}{Longfei Wu}, \bibinfo{person}{Guisheng Yin}, \bibinfo{person}{Lijie Li}, {and} \bibinfo{person}{Hongbin Zhao}.} \bibinfo{year}{2017}\natexlab{}.
\newblock \showarticletitle{A Survey on Security and Privacy Issues in Internet-of-Things}.
\newblock \bibinfo{journal}{\emph{IEEE Internet of Things Journal}} \bibinfo{volume}{4}, \bibinfo{number}{5} (\bibinfo{year}{2017}), \bibinfo{pages}{1250--1258}.
\newblock
\urldef\tempurl%
\url{https://doi.org/10.1109/JIOT.2017.2694844}
\showDOI{\tempurl}


\bibitem[Yang et~al\mbox{.}(2013)]%
        {yang2013rssi}
\bibfield{author}{\bibinfo{person}{Zheng Yang}, \bibinfo{person}{Zimu Zhou}, {and} \bibinfo{person}{Yunhao Liu}.} \bibinfo{year}{2013}\natexlab{}.
\newblock \showarticletitle{From RSSI to CSI: Indoor localization via channel response}.
\newblock \bibinfo{journal}{\emph{ACM Computing Surveys (CSUR)}} \bibinfo{volume}{46}, \bibinfo{number}{2} (\bibinfo{year}{2013}), \bibinfo{pages}{1--32}.
\newblock


\bibitem[Yeh et~al\mbox{.}(2020)]%
        {yeh2020energy}
\bibfield{author}{\bibinfo{person}{Ling-Yu Yeh}, \bibinfo{person}{Po-Jen Chen}, \bibinfo{person}{Chen-Chun Pai}, {and} \bibinfo{person}{Tsung-Te Liu}.} \bibinfo{year}{2020}\natexlab{}.
\newblock \showarticletitle{{An energy-efficient dual-field elliptic curve cryptography processor for Internet of Things applications}}.
\newblock \bibinfo{journal}{\emph{IEEE Transactions on Circuits and Systems II: Express Briefs}} \bibinfo{volume}{67}, \bibinfo{number}{9} (\bibinfo{year}{2020}), \bibinfo{pages}{1614--1618}.
\newblock


\bibitem[Zhang et~al\mbox{.}(2017)]%
        {Zhang:2017}
\bibfield{author}{\bibinfo{person}{Jiansong Zhang}, \bibinfo{person}{Zeyu Wang}, \bibinfo{person}{Zhice Yang}, {and} \bibinfo{person}{Qian Zhang}.} \bibinfo{year}{2017}\natexlab{}.
\newblock \showarticletitle{{Proximity based IoT device authentication}}. In \bibinfo{booktitle}{\emph{IEEE INFOCOM}}.
\newblock


\bibitem[Zhu and Alsharari(2015)]%
        {zhu2015improved}
\bibfield{author}{\bibinfo{person}{Haiping Zhu} {and} \bibinfo{person}{Talal Alsharari}.} \bibinfo{year}{2015}\natexlab{}.
\newblock \showarticletitle{An improved RSSI-based positioning method using sector transmission model and distance optimization technique}.
\newblock \bibinfo{journal}{\emph{International Journal of Distributed Sensor Networks}} \bibinfo{volume}{11}, \bibinfo{number}{9} (\bibinfo{year}{2015}), \bibinfo{pages}{587195}.
\newblock


\end{thebibliography}

\end{document}